\documentclass[11pt]{article}
\pdfoutput=1
\usepackage{amsmath,amssymb,epsfig,amsfonts}
\usepackage{graphicx,subfigure}
\usepackage[usenames, dvipsnames]{color}
\usepackage[pagebackref]{hyperref}
\usepackage{cite}
\usepackage{verbatim} 
\usepackage{float}
\restylefloat{table}
\usepackage[all]{xy}
\usepackage{pdflscape}
\usepackage{tikz}
\usepackage{array}
\usepackage{youngtab}
\usepackage{multirow}
\usepackage{color}
\usepackage{ulem}
\usepackage{tikz-cd}

\DeclareFontFamily{U}{wncy}{}
    \DeclareFontShape{U}{wncy}{m}{n}{<->wncyr10}{}
    \DeclareSymbolFont{mcy}{U}{wncy}{m}{n}
    \DeclareMathSymbol{\Sh}{\mathord}{mcy}{"58} 

\makeatletter

\renewcommand*{\backref}[1]{}
\renewcommand*{\backrefalt}[4]{({%
    \ifcase #1 Not cited.%
          \or Page~#2.%
          \else Pages #2.%
    \fi%
    })}

\def\CO{{\mathcal{O}}}
\def\CE{{\mathcal{E}}}

\def\bbP{{\mathbb{P}}}
\def\bbZ{{\mathbb{Z}}}

\def\fkg{{\mathfrak{g}}}

\newcount\hour \newcount\minute
\hour=\time \divide \hour by 60
\minute=\time
\def\now{%
\ifnum \hour<13
  \ifnum \hour=0 \advance \hour by 12 \number\hour:\else \number\hour:\fi%
     \ifnum \minute<10 0\fi%
     \number\minute%
\ A.M.%
\else \advance \hour by -12 \number\hour:%
  \ifnum \minute<10 0\fi%
  \number\minute%
  \ P.M.%
\fi%
}

\makeatother
\hypersetup{
    pdftitle={When Rational Sections Become Cycle: Gauge Enhancement in F-theory via Mordell--Weil Torsion},
    pdfauthor={Florent Baume, Mirjam Cvetic, Craig Lawrie, Ling Lin},
    pdfsubject={F-theory compactifications, Mordell--Weil torsion, genus-one fibration}
}

\begin{document}

\baselineskip=18pt  
\numberwithin{equation}{section}  
\allowdisplaybreaks  



\vspace*{-2cm} 
\begin{flushright}
{\tt UPR-1287-T}\\
\end{flushright}

\vspace*{0.8cm} 
\begin{center}
 {\LARGE  When Rational Sections Become Cyclic}\\
 \vspace*{0.2cm}
 {\Large Gauge Enhancement in F-theory via Mordell--Weil Torsion}

 \vspace*{1.8cm}
 {Florent Baume$\,^1$, \, Mirjam Cveti\v{c}$\,^{2,3}$, \, Craig Lawrie$\,^1$, \, Ling Lin$\,^2$}\\

 \vspace*{1.2cm} 
{\it $^1$ Institut f\"ur Theoretische Physik, Ruprecht-Karls-Universit\"at,\\
 Philosophenweg 19, 69120 Heidelberg, Germany }\\
  
\bigskip
{\it $^2$ Department of Physics and Astronomy, University of Pennsylvania, 209
  S.~33rd Street, \\
 Philadelphia, PA 19104-6396, USA}\\

 \bigskip
{\it $^3$ Center for Applied Mathematics and Theoretical Physics, \\
 University of Maribor, Maribor, Slovenia}\\

 \bigskip
  
 {\tt f.baume@thphys.uni-heidelberg.de}$\,,\quad$ {\tt cvetic@physics.upenn.edu}, {\tt
   {gmail:$\,$craig.lawrie1729}}$\,,\quad$ {\tt lling@physics.upenn.edu}

%
%
%
\vspace*{0.8cm}
\end{center}
\vspace*{.5cm}
%
\noindent
We explore novel gauge enhancements from abelian to non-simply-connected gauge
groups in F-theory. To this end we consider complex structure deformations of
elliptic fibrations with a Mordell--Weil group of rank one and identify the
conditions under which the generating section becomes torsional.  For the
specific case of $\mathbb{Z}_2$ torsion we construct the generic solution to
these conditions and show that the associated F-theory compactification
exhibits the global gauge group $[SU(2) \times SU(4)]/\mathbb{Z}_2 \times
SU(2)$. The subsolution with gauge group $SU(2) / \mathbb{Z}_2 \times SU(2)$,
for which we provide a global resolution, is related by a further complex
structure deformation to a genus-one fibration with a bisection whose Jacobian
has a $\mathbb{Z}_2$ torsional section. While an analysis of the spectrum on
the Jacobian fibration reveals an $SU(2)/\mathbb{Z}_2 \times \mathbb{Z}_2$
gauge theory, reproducing this result from the bisection geometry
raises some conceptual puzzles about F-theory on genus-one fibrations.

\newpage

\tableofcontents
\newpage

\section{Introduction}\label{sec:intro}

\subsection{Motivation and Summary}

It has been well-established that F-theory
\cite{Vafa:1996xn,Morrison:1996na,Morrison:1996pp} is an efficacious tool for
the geometric engineering of non-perturbative string theory vacua in even
dimensions.  An F-theory vacuum is associated to an elliptically fibered
Calabi--Yau variety, $\pi: Y \rightarrow B$, where the complex structure of
the torus above each point of $B$ specifies the value of the axio-dilaton at
that point in a Type IIB compactification on $B$. The requirement that the
axio-dilaton value at each point can be glued together to form a global
elliptic fibration is necessary for the consistency of such a vacuum. Any
elliptic fibration \cite{MR0387292} can be cast in the form of a Weierstrass
equation
\begin{equation}\label{eqn:WSintro}
  y^2 = x^3 + f \, x\, z^4 + g \, z^6 \,,
\end{equation}
where $[x:y:z]$ are the coordinates of an ambient $\bbP_{231}$.
The condition that this elliptic
fibration is Calabi--Yau is expressed through the bundles that the
coefficients $f$ and $g$ are global sections of, 
\begin{equation}
  f \in H^0(B, \mathcal{O}(-4K_B)) \,,\qquad g \in H^0(B, \mathcal{O}(-6K_B)) \,,
\end{equation}
with $K_B$ the canonical class of $B$. One significant sector of the geometric
engineering of F-theory vacua involves finding necessary and sufficient
conditions on the coefficients $f$ and $g$ such that the vacuum has particular
physical features. 
For instance, the vanishing orders of $f$ and $g$ (and $\Delta = 4 f^3 + 27
g^2$) along certain divisors in $B$ indicate the presence of singular fibers,
by the Kodaira--N\'eron classification \cite{MR0132556,MR0184257,MR0179172},
which are in direct correspondence with the
non-abelian gauge algebra in the F-theory effective physics.

In this paper we are interested in studying under what circumstances the
non-abelian gauge group, $G$, in the low-energy physics has a non-trivial
fundamental group. As is known from \cite{Aspinwall:1998xj, Aspinwall:2000kf, Mayrhofer:2014opa}, 
$\pi_1(G)$ is associated to the existence of torsion inside of the
Mordell--Weil group of the elliptic fibration. The Mordell--Weil group is the
group of rational sections of the elliptic fibration, which is a finitely
generated abelian group \cite{mordell1922rational}, 
\begin{equation}
  \text{MW}(Y) \cong \mathbb{Z}^r \times \Gamma \,,
\end{equation}
and the group structure comes from the
fiberwise application of the elliptic curve group law. The free
part, $\bbZ^r$, of the Mordell--Weil group gives rise to a $U(1)^{r}$ symmetry
in F-theory \cite{Morrison:1996na, Morrison:1996pp}, and the torsion part,
$\Gamma$, is
related to the global structure of the non-abelian gauge group,\footnote{Note that such a global structure arising from torsional sections can never affect the abelian part of the gauge group.
Instead, non-trivial gauge group structures including abelian symmetries arise directly from rational sections generating the free part of the Mordell--Weil group \cite{Cvetic:2017epq}.}
\begin{equation}
  \pi_1(G) \cong \Gamma \,.
\end{equation}

In order to determine the requirements for the existence of torsion inside of
the Mordell--Weil group, we shall explore deformations of geometries such that
a free section becomes torsional.  In other words, we are considering a family
of elliptic fibrations with Mordell--Weil rank $r+1$ without torsion, and
whose central fiber is a fibration of MW-rank $r$ and non-trivial torsion
$\Gamma = \mathbb{Z}_n$.  Without loss of generality one can restrict one's attention to $r =
0$, and we shall do so henceforth.  Physically, the complex
structure deformation from the generic to the central fiber of the family
corresponds to the enhancement of a $U(1)$ into a non-simply-connected gauge
group with $\pi_1(G) \cong \Gamma$.  Indeed, the enhancement of $U(1)$s into
non-abelian symmetries has been studied extensively in recent literature.
There, the process involves tuning two rational sections to sit atop one
another.  This has proven to be very effective in constructing global F-theory
compactifications with higher dimensional representations, most recently
$SU(3)$ models with a $\bf 6$ representation \cite{Cvetic:2015ioa} and $SU(2)$
models with $\bf 4$ representation \cite{Klevers:2016jsz}.  

In this article we provide another approach to enhancement by 
tuning a rational section not to collide with another, but to sit globally at a
specific $\Gamma$ torsional point of the elliptic fiber. 
A special case of such a tuning was discussed in \cite{Mayrhofer:2014opa}. 
This paper puts the idea on a general footing and, by doing so, extends the
network of known Higgsing chains of F-theory compactifications.  Whether it is
possible to construct examples of previously unconstructed representations for
F-theory matter in this manner of tuning remains an open question.

The generic elliptic fibration with Mordell--Weil torsion can be
generated by the following process.
\begin{enumerate}
  \item One begins with an elliptic fibration with a rank one torsion-free
    Mordell--Weil group. It has been shown in \cite{Morrison:2012ei} that any
    elliptic fibration with a rank one Mordell--Weil group is birationally
    equivalent to a Weierstrass model (\ref{eqn:WSintro}) with specific forms of $f$ and $g$.
  \item Use the group law on the elliptic curve for this Weierstrass
    equation to determine the $\Gamma$ torsional points, as described in
    appendix \ref{sec:MW-torsion_Weierstrass}, with rational coordinates
    \begin{equation}\label{eqn:Gtors}
	  [x_\Gamma : y_\Gamma : z_\Gamma] \,.
	\end{equation} 
  \item Let $[x_Q : y_Q :z_Q]$ be the coordinates of the free rational section
    in each fiber for the rank one model obtained in step 1, and then solve
    \begin{equation}
      [x_Q : y_Q : z_Q] = [x_\Gamma : y_\Gamma : z_\Gamma] \,,
    \end{equation}
    as a global polynomial relation inside the function field of the base of the
    fibration.
\end{enumerate}
At this point one has determined the sufficient and necessary conditions on
$f$ and $g$ such that the elliptic fibration has Mordell--Weil group $\Gamma$.
It now remains to study the physics associated to this generic model, such as
the non-abelian gauge group and how $\pi_1(G)$ acts if $G$ is not
simple.
However, one subtlety quickly arises; while every elliptic fibration with a
rank one Mordell--Weil group is birationally equivalent to the generic model
written down in \cite{Morrison:2012ei}, the birational transformation does not
necessarily preserve the canonical class.  Thus the elliptic fibration that is
constructed in step 1 above may not be Calabi--Yau, and hence may not be
immediately amenable to F-theory.  The effective physics in
\cite{Morrison:2012ei}, and in most of the subsequent literature, for example
\cite{Mayrhofer:2012zy,Braun:2013yti,Cvetic:2013nia,Braun:2013nqa,Borchmann:2013hta,Morrison:2014era,Antoniadis:2014jma,Kuntzler:2014ila,Esole:2014dea,Cvetic:2015uwu,Wang:2016urs},
assumed that the generic model was itself Calabi--Yau, however recent work
\cite{Klevers:2014bqa,Cvetic:2015moa,Lawrie:2015hia,Krippendorf:2015kta,Cvetic:2015ioa,Klevers:2016jsz,Morrison:2016xkb}
has begun to explore the physics of those models where the generic model is
not Calabi--Yau. The difference between the two situations is controlled by
the height, or ``tallness'' \cite{Morrison:2016xkb}, of the rational section,
which we review in section \ref{sec:rev}.

In this paper we compute the generic form of an elliptic fibration with
$\mathbb{Z}_2$ torsion that arises from a complex structure deformation of the
general elliptic fibration with rank one Mordell--Weil group. Finding the most general
solution to the torsional section condition is done in section
\ref{sec:Z2enc}. Specializing the $\bbZ_2$ torsion model to a Calabi--Yau
threefold, the physics of the F-theory compactification is explored in
section \ref{sec:F-theory_on_Z2}, making use of the structure of the tuned
Weierstrass model and the anomaly constraints \cite{Park:2011ji} on the
resulting 6D $\mathcal{N} = (1,0)$ supergravity theory. In section
\ref{sec:resolution_restricted_model} we construct an explicit resolution of
singularities of a singular elliptic threefold which realizes a subsolution of
the generic $\mathbb{Z}_2$ torsional section condition and observe the
torsional section explicitly. Furthermore the case of $\Gamma = \mathbb{Z}_3$
is discussed in appendix \ref{sec:Z3_unhiggsing}. 

It is known in examples that there is an intimate relationship between the
torsion subgroup of an elliptic fibration and the multi-section geometry 
of a certain ``dual'' genus-one fibration \cite{Klevers:2014bqa},
where this notion of duality has mainly been explored when the elliptic
fibration can be written as a toric complete intersection \cite{Oehlmann:2016wsb, Cvetic:2016ner}.
Where it has been studied such a mapping exchanges the Tate--Shafarevich
group, which is the group of genus-one fibrations with the same Jacobian fibration and no isolated multiple
fibers, and the torsion subgroup of the
Mordell--Weil group. In section \ref{sec:rosie} we apply the understanding of
$\mathbb{Z}_2$ Mordell--Weil torsion acquired in the rest of the paper to
construct a non-toric genus-one fibration with a bisection whose Jacobian elliptic fibration 
has Mordell--Weil group $\mathbb{Z}_2$. This example raises important questions
about F-theory on genus-one fibrations.

\subsection{Review of Elliptic Fibrations with Rank One Mordell--Weil Group}\label{sec:rev}

In \cite{Morrison:2012ei} a general form for an elliptic fibration with Mordell--Weil rank one was written down.  This model,
occasionally referred to as the Morrison--Park model, is given by a
Weierstrass form (\ref{eqn:WSintro}) with specialized coefficients, $f$ and $g$:
\begin{equation}\label{eqn:U1WS}
  y^2 = x^3 + \left( c_1 c_3 - \frac{1}{3} c_2^2 - b^2 c_0 \right) x \, z^4 + \left( -c_0 c_3^2
  + \frac{1}{3} c_1 c_2 c_3 - \frac{2}{27} c_2^3 +
        \frac{2}{3} b^2 c_0 c_2 - \frac{1}{4} b^2 c_1^2 \right) z^6 \,.
\end{equation}
The coordinates $[x : y : z]$ are again the projective coordinates in the fiber of
the $\mathbb{P}_{231}$ fibration over $B$ in which this equation cuts out a
hypersurface. 
The coefficients $b$ and $c_i$ are sections of certain line
bundles over the base of the fibration. 
While any elliptic fibration, of any dimension, with Mordell--Weil rank one is birationally equivalent to \eqref{eqn:U1WS}, there is no requirement that the
canonical class will be preserved under this birational map. Thus one could
begin with a Calabi--Yau elliptic fibration and find a birationally equivalent
elliptic fibration of the form (\ref{eqn:U1WS}) which is not Calabi--Yau.  We
write the Weierstrass line bundle \cite{MR1078016}, of which $f$ and $g$
transform as sections of the fourth and sixth powers, of the resulting elliptic
fibration (\ref{eqn:U1WS}) as\footnote{We choose this form for the Weierstrass line bundle to be able to write the Calabi--Yau condition
for the elliptic fibration as $D = 0$.}
\begin{equation}
  \mathcal{O}(-K_B + D) \,,
\end{equation}
where $K_B$ is the canonical class of $B$ and $D$ is a divisor class on $B$. Up
to a choice of twisting bundle $\mathcal{O}(\beta)$, the classes associated to
the specialized coefficients can be determined through the classes of $f$ and $g$.
First, it can easily be seen that the class of $c_0$ must be even, and so it
can be fixed via
\begin{equation}
  [c_0] = 2\beta \,,
\end{equation}
and thus the rest of the classes follow, giving
\begin{align}\label{tab:coefficients_classes_MP}
	\begin{array}{c|c|c|c|c|c}
    \text{Coefficient} & b & c_0 & c_1 & c_2 & c_3 \\ \hline \rule{0pt}{2.5ex}
    \text{Class} & - 2 K_B + 2 D - \beta & 2 \beta & -K_B + D + \beta & - 2 K_B + 2 D & - 3 K_B + 3 D - \beta 
	\end{array} \,.
\end{align}
Because the coefficients have to be globally well-defined sections of corresponding line bundles, the classes in \eqref{tab:coefficients_classes_MP} must not be anti-effective.
This constraints the possible choices of classes $\beta$ and $D$ for a given base $B$.

Given the Weierstrass model (\ref{eqn:U1WS}) it is easy to see that there are generically two
independent rational sections located at
\begin{equation}\label{eqn:WSsecs}
  \begin{aligned}
    S_0 &: & \quad [x_0 :  y_0 : z_0] &= [1 : 1 : 0] \,,\cr
    S_Q &: & \quad \left[x_Q : y_Q : z_Q \right] &= \left[ c_3^2 - \frac{2}{3}
      b^2 \, c_2 \, : \, -c_3^3 + b^2 \, c_2 \, c_3 - \frac{1}{2} \, b^4 \,
    c_1 \, : \, b\right] \,. 
  \end{aligned}
\end{equation}
We note that the generic Weierstrass model in (\ref{eqn:U1WS}) is singular, in
particular the fiber above the locus $b = c_3 = 0$ is a nodal rational
curve. One can see the singular nature of the Weierstrass model by observing
the discriminant of (\ref{eqn:U1WS}):
\begin{equation}\label{eqn:Disc}
  \begin{aligned}
   \Delta &= 64 b^6 c_0^3 +16 c_0^2 \left(8 b^4 c_2^2+12 b^4 c_1 c_3+36 b^2
   c_2 c_3^2+27 c_3^4\right)
   \cr &\quad -8 c_0 \left(-8 c_2^3 \left(b^2 c_2+c_3^2\right)+4 c_1 c_3 c_2 \left(10
      b^2 c_2+9 c_3^2\right)+3 b^2 c_1^2 \left(6 b^2 c_2+c_3^2\right)\right)
      \cr &\quad +c_1^2 \left(27 b^4 c_1^2-16 c_2^2 \left(b^2
      c_2+c_3^2\right)+8 c_1 \left(9 b^2 c_2 c_3+8 c_3^3\right)\right) \,.
  \end{aligned}
\end{equation}

An alternate way to realize the elliptic fibration (\ref{eqn:U1WS}) is via a
birationally equivalent, singular, hypersurface in a $\mathbb{P}_{112}$
fibration over $B$:
\begin{equation}\label{eq:MP-fibration}
  w^2 + b \,v^2\,w = u\,(c_0 \, u^3 + c_1 \, u^2 \, v + c_2\, u\, v^2 + c_3\,v^3) \, .
\end{equation}
Here $[u : v : w]$ are the coordinates on the $\mathbb{P}_{112}$, and the
coefficients are the same sections as in (\ref{tab:coefficients_classes_MP}). The two
rational sections are now exhibited clearly: setting $u = 0$ yields the
reducible equation $w^2 + b \, v^2\,w = 0$. Therefore the two points
\begin{equation}\label{eqn:Psecs}
  [0 : 1 : 0] \quad \text{and} \quad [0 : 1 : -b ] \,,
\end{equation}
in the ambient $\mathbb{P}_{112}$ fiber mark two distinct points on each
elliptic fiber, and thus give rise to two distinct rational sections.  Indeed such a
hypersurface was determined in \cite{Morrison:2012ei} to be the generic form
for elliptic fibrations with two rational sections, using analogous arguments
to those in \cite{MR0387292}
for deriving the generic Weierstrass equation for elliptic fibrations, which
have one rational section. Such a construction can then be mapped into the
specialized Weierstrass form (\ref{eqn:U1WS}) using Nagell's algorithm
\cite{MR1555271}.

As discussed, for generic choices of coefficients $b, c_i$, the section $S_Q$
generates a rank one Mordell--Weil (sub-)group. F-theory compactified on this
space, or more precisely on a Calabi--Yau space in the same birational
equivalence class, therefore has a $U(1)$ gauge factor. The massless spectrum of
this theory contains hypermultiplets which are charged under the $U(1)$ gauge
group, and the range of charges that arise generically depends on the divisor
class $D$ \cite{Morrison:2016xkb}. 

The requirement for the existence of certain charged states depending on $D$
can be seen through the N\'eron--Tate height.
The height of a rational section is the projection to $B$ of the
self-intersection of the divisor associated to the Shioda map, $\sigma(S_Q)$, of the section
\cite{MR1030197,MR1081832}:
\begin{equation}
  h(S_Q) = - \pi(\sigma(S_Q), \sigma(S_Q)) \,.
\end{equation}
This height, which is a divisor in $B$, is related to the anomaly of the
$U(1)$ gauge factor associated to the section, if $B$ is a twofold base, and
anomaly cancellation further relates the charges of the $U(1)$ charged
hypermultiplets to $h(S_Q)$ \cite{Park:2011ji}. Assuming that the
model (\ref{eqn:U1WS}) has no codimension-one singularities except the generic
type $I_1$ fibers, the height was worked out in \cite{Morrison:2012ei}, and
further one can see that it is bounded by the requirement that $\beta$ be an
effective divisor class,
\begin{equation}\label{eqn:hineq}
  h(S_Q) \leq - 6 K_B + 4 D \,.
\end{equation}
From the height one can define a notion of so-called
tallness \cite{Morrison:2016xkb}, which is
constrained by the inequality (\ref{eqn:hineq}),
\begin{equation}
  t(S_Q) \equiv \frac{h(S_Q) \cdot h(S_Q)}{-2K_B \cdot h(S_Q)} \leq \max_I
  q_I^2 \,,
\end{equation}
where the product is understood as the intersection product on the twofold base $B$.
Here the index $I$ runs over the charged hypermultiplets, which have charge
$q_I$. It is found that the value of $t(S_Q)$ fixes a minimal largest charge
that is forced to appear in the model for a consistent, anomaly free theory.

We will be principally interested in computing the spectrum in the case (\ref{eqn:U1WS}) itself defines a Calabi--Yau
elliptic fibration, in which case $D = 0$, and the bound (\ref{eqn:hineq}) is
saturated when
\begin{equation}
  h(S_Q) = - 6 K_B \,,
\end{equation}
for which the tallness of the section is
\begin{equation}
  t(S_Q) = 2 \,.
\end{equation}
As such, the theory is required to contain a hypermultiplet of charge at least
$2$, but not necessarily of any higher charge.
Indeed if one studies the generic model (\ref{eqn:U1WS}) then one can observe
that the matter spectrum of this theory, with $D = 0$, consists of charge $1$ and $2$
hypermultiplets \cite{Morrison:2012ei}. In \cite{Klevers:2014bqa} it was argued that given any Weierstrass
model with non-trivial Mordell--Weil generator $S_Q$, the union of all singlet
matter loci is given by the complete intersection 
\begin{equation}
  V(y_Q, 3\,x_Q^2 + f\,z_Q^4) = V(\partial_y {P_W}|_Q, \partial_x P_W|_Q) \,.
\end{equation}
For the generic single $U(1)$ model \eqref{eqn:U1WS} that we consider, this is in
agreement with the results in \cite{Morrison:2012ei}:
\begin{align}
		V(I) := V(y_Q, 3\,x_Q^2 + f\,z_Q^4) = 
		\left(
		\begin{array}{r}
			\displaystyle -c_3^3 + b^2\, c_2\, c_3 - \frac{1}{2}\, b^4\, c_1 = 0 \\ \rule{0pt}{3ex}
			c_3^4 - 2\,b^2 \, c_2 \, c_3^2 + b^4\,c_2^2 - b^6\,c_0 = 0
		\end{array}
		\right)  \,.
		\label{eq:singlet_loci_MP}
\end{align}
The ideal $I$ has two associated primes, $\mathfrak{p}_1$ and $\mathfrak{p}_2
= (b, c_3)$, with the charge $2$ singlets localized at $V(\mathfrak{p}_2)$,
and the charge $1$ singlets sit at $V(\mathfrak{p}_1)$. It can be shown that in terms of cycle classes, we
have 
\begin{equation}
  [V(I)] = [V(\mathfrak{p}_1)] + 16\,[V(\mathfrak{p}_2)] \,.
\end{equation}
Thus, the multiplicities of the charged singlets are given in terms of intersection numbers as:
\begin{align}\label{multMP}
	\begin{split}
    \text{Charge 1:}\qquad& x_1 = [V(\mathfrak{p}_1)] = [c_1^4] \cdot [c_3]
    =  12 \, K_B^2 - 8\, K_B\cdot\beta-4 \, \beta^2 \, , \\ 
		\text{Charge 2:}\qquad& x_2 = [V(\mathfrak{p}_2)] = [b]\cdot [c_3] = 6 \, K_B^2 + 5\,K_B\cdot\beta+\beta^2 \, .
	\end{split}
\end{align}

\section[Gauge Enhancement via \texorpdfstring{\boldmath{$\mathbb{Z}_2$}}{Z2} Torsion]{Gauge Enhancement via \boldmath{$\mathbb{Z}_2$} Torsion}\label{sec:Z2enh}

In this section we will determine a Weierstrass fibration that is birationally
equivalent to any elliptic fibration with Mordell--Weil torsion $\Gamma =
\mathbb{Z}_2$. For the construction we assume, as discussed in section
\ref{sec:intro}, that the elliptic fibration fits into a family of elliptic
fibrations with a $U(1)$ gauge group, however we do not require any
constraints on the dimension of the elliptic fibration, or on its canonical
class.

We begin by utilizing that a generic element of such a family of elliptic
fibrations is birationally equivalent to a Weierstrass equation of the form
(\ref{eqn:U1WS}). If the rational section, $S_Q$, located at the point
(\ref{eqn:WSsecs}) in the Weierstrass model is to be situated globally at the
$\mathbb{Z}_2$ torsion point of the elliptic fiber, then one must, as has been
determined in appendix \ref{sec:MW-torsion_Weierstrass}, satisfy
\begin{equation}\label{eqn:Z2tunecond}
  y_Q =  -c_3^3 + b^2 c_2 c_3 - \frac{1}{2} b^4 c_1 = 0 
\end{equation}
as a globally valid equation.

From the ideal \eqref{eq:singlet_loci_MP} giving the codimension two loci in
the base at which the degenerate fibers, and thus the matter hypermultiplets,
are located in the $U(1)$ model we can see that one of the two generators is
$y_Q$. As such, it is evident that, after solving (\ref{eqn:Z2tunecond}) the
second equation of the ideal, 
\begin{equation}
  3x_Q^2 + fz_Q^4=0 \,,
\end{equation}
defines a codimension one locus of degenerate fibers, which will in turn give
rise to a non-abelian gauge algebra. We point out that the
compensation for the loss of a $U(1)$ gauge group by a non-abelian gauge
group, $G$, is expected as the F-theory gauge algebra must have a center which
contains a $\mathbb{Z}_2$ such that it is consistent to have $\pi_1(G) =
\mathbb{Z}_2$.  In the following, we will explicitly determine the non-abelian
gauge group of the enhanced theory, which turns out to be more intricate
than, perhaps, naively expected.

\subsection[Deforming to \texorpdfstring{$\mathbb{Z}_2$}{Z2} Torsion]{Deforming to \boldmath{$\mathbb{Z}_2$} Torsion}\label{sec:Z2enc}

To solve the tuning condition $y_Q = 0$ as a globally valid equation, we
examine $y_Q$ -- which is a global section of some line bundle -- locally,
through the restriction of $y_Q$ to local rings of function germs
$\CO_{B,p}$.  The assumption of a smooth base $B$ implies that for any point
$p \in B$, this local ring is a unique factorization domain (UFD)
\cite{MR0103906}.  Intuitively, one can think about a UFD as a polynomial
ring, in which every polynomial can be factorized uniquely (up to units,
i.e., constant pre-factors) into powers of prime, or irreducible, elements.
Over a UFD, the equation $y_Q = 0$ can be solved by systematically by
determining common factors $r_i$ of individual terms until the equation can be
linearly
satisfied.  For the solution to be valid globally, one has to check
that there exist appropriate line bundles with global sections that restrict to
$r_i$ at the local rings $\CO_{B,p}$.

For the case at hand, where we wish to solve (\ref{eqn:Z2tunecond}), we observe that
\begin{equation}\label{eq:tuning_condition}
	y_Q = 0 \quad \Leftrightarrow \quad  b^2 \left( c_2\,c_3 -
  \frac{b^2\,c_1}{2} \right) = c_3^3 \,,
\end{equation}
which implies that $b$ must divide $c_3^3$. One can begin the process of solving
this particular global polynomial equation over the unique factorization
domain by writing the coordinates $b$ and $c_3$ in terms of their coprime
decomposition
\begin{equation}
  b = \sigma \, s \,, \quad c_3 = \sigma \, r \,,
\end{equation}
where now $r$ and $s$ are coprime over the UFD. By direct substitution the
polynomial (\ref{eq:tuning_condition}) becomes
\begin{equation}\label{eqn:restuning}
  \sigma^3\, (s^2 \, (c_2 \, r - \sigma \, s^2 \, c_1 / 2) - r^3) = 0 \,.
\end{equation}
The first solution of the tuning condition that would give rise to a torsional
section would be if $\sigma$ vanished globally, however in such an eventuality
one can see that the discriminant, given in (\ref{eqn:Disc}), also vanishes
globally, and thus the putative elliptic fibration is everywhere degenerate.
We must thus only consider solutions with the vanishing of the second factor
in (\ref{eqn:restuning}). The form of the second factor requires $s^2$ to divide $r^3$, however since $r$ and $s$ are coprime this is only possible
if $s$ is globally constant. As such the first requirement that
(\ref{eq:tuning_condition}) holds as a global equation is that
\begin{equation}
  c_3 = b \, r \,,
\end{equation}
and the remnant equation that must be solved is
\begin{equation}
  r \, (c_2 - r^2) - \frac{1}{2} \, b\, c_1 = 0 \,.
\end{equation}
Solving this equation generically over a unique factorization domain yields
the solution (see the appendices of \cite{Kuntzler:2014ila,Lawrie:2014uya} for details)
\begin{equation}\label{eq:tuning_generic_solution}
  b = s_1 s_3 \,, \quad c_1 = 2 s_2 s_4 \,, \quad c_2 = s_3 s_4 + s_1^2 s_2^2
  \,, \quad
  c_3 = s_1^2 s_2 s_3 \,,
\end{equation}
where $(s_1, s_4)$ and $(s_2, s_3)$ are coprime pairs.

The solution is generic in that this solution is the most general way to solve
the equation $(\ref{eq:tuning_condition})$ over a UFD. The classes of the
$s_i$ are determined by \eqref{eq:tuning_generic_solution} and the classes \eqref{tab:coefficients_classes_MP}. They can be
expressed in terms of $K_B$, $\beta$, $D$, and a further, a priori arbitrary, class
$\Sigma$:
\begin{align}\label{eq:tuning_solution_classes}
  [s_1] = \Sigma \, , \quad [s_2] = -K_B+D-\Sigma \, , \quad [s_3] = -2K_B +2D -
  \Sigma - \beta \, , \quad [s_4] = \Sigma + \beta \, .
\end{align}
These classes must not be anti-effective in order for the $s_i$ to be globally
well-defined. For a fixed choice of base $B$, $\beta$, and $D$ this requirement
constrains $\Sigma$ in terms of $K_B$, $\beta$, and $D$. 

For example, if $B =
\bbP^2$, and thus $-K_B = 3H$, where $H$ is the hyperplane class, and
we further  choose $\beta = n\,H$, $D = 0$, and $\Sigma = k\,H$
then the effectiveness requirement is satisfied if either
\begin{equation}
  \begin{aligned}
	0 \leq n \leq 3 \quad & \Longrightarrow \quad 0 \leq k \leq 3 \, , \cr
	3 < n < 6 \quad & \Longrightarrow \quad 0 \leq k \leq 6-n \, ,\cr
	n = 6 \quad & \Longrightarrow \quad k = 0 \, .
  \end{aligned}
\end{equation}
Note that the range of $n$ is dictated by the effectiveness of the classes in
(\ref{tab:coefficients_classes_MP}).

There are a multitude of specialized solutions for the tuning of a
$\mathbb{Z}_2$ torsional section that arise when the generic solution
(\ref{eq:tuning_generic_solution}) is applied with non-generic $s_i$. One
relevant specialized tuning, which will be explored in more detail in section
\ref{sec:resolution_restricted_model}, is 
\begin{equation}
  s_2 = 0 \,.
\end{equation}
After such a tuning one can see that the coefficients $b$ and $c_2$ are now
simply written in terms of their coprime decomposition, with common factor
$s_3$. 
One can in addition seek the constraint that $b$ and $c_2$ be generic divisors
in $B$, which requires that the intersection of the two divisors be in
codimension $\geq 2$; this implies that there is no common component, or that
$s_3$ does not vanish anywhere along $B$. For $s_3$ to be a constant function
on $B$ it is necessary that it transform as a section of $\mathcal{O}_B$.
This is fixed, in addition to $s_1$ and $s_4$ being of the
same classes as, respectively, $b$ and $c_2$, by imposing
\begin{equation}
  \Sigma = -2 {K}_B - \beta \,. 
\end{equation}
This particular specialization, whose explicit resolution will be studied in
section \ref{sec:resolution_restricted_model}, can be said to correspond to
the generic solution (\ref{eq:tuning_generic_solution}) with the additional
conditions that
\begin{equation}
  s_2 = 0 \,, \quad s_3 = 1 \,.
\end{equation}
As such this specialized solution merely corresponds to setting
\begin{equation}
  c_1 = 0 \,, \quad c_3 = 0 \,,
\end{equation}
with $b$, $c_2$, and $c_0$ generic. 
There are many further specializations which change the structure and
configuration of the singular fibers in the elliptic fibration whilst
retaining the required $\mathbb{Z}_2$ torsional section, however as they are
all specialized solutions of (\ref{eq:tuning_generic_solution}) they shall not be
explicitly considered further here.

\subsection[F-theory of the \texorpdfstring{$\bbZ_2$}{Z2} Torsional
Model]{F-theory of the \boldmath{$\bbZ_2$} Torsional Model}\label{sec:F-theory_on_Z2}

In this section we elaborate on some of the effective physics of an
F-theory compactification on the generic elliptic fibration with
$\mathbb{Z}_2$ Mordell--Weil torsion, as given through the Weierstrass
elliptic fibration (\ref{eqn:U1WS}) with (\ref{eq:tuning_generic_solution}).
In this we are hampered if the elliptic fibration has non-trivial canonical
class, and thus for simplicity, we consider F-theory compactifications to 6D
on elliptic Calabi--Yau threefolds of the specified form. The restriction to Calabi--Yau
is equivalent to taking the divisor class $D$ to be trivial in
(\ref{tab:coefficients_classes_MP}) and (\ref{eq:tuning_solution_classes}).
The advantage of 6D compactifications is that there are strong anomaly
conditions \cite{Park:2011ji} with which we can bootstrap the spectrum without
an explicit resolution. 

If we plug the generic solution \eqref{eq:tuning_generic_solution} into the
expressions for $f$ and $g$ in \eqref{eqn:U1WS}, we obtain:
\begin{equation}\label{eq:Weierstrass_functions_generic_solution}
  \begin{aligned}
		f &= -c_0s_1^2s_3^2 + 2s_1^2s_2^2s_3s_4 - 
      \frac{1}{3}(s_1^2s_2^2 + s_3s_4)^2 \,, \cr
		g &= \frac{1}{27}\, (s_ 1^2\, s_ 2^2 - 2\, s_ 3\, s_ 4)\, (2\, s_ 1^4\, s_
    2^4 - s_ 3^2\, s_ 4^2 + s_ 1^2\, s_ 3\, (9\, c_ 0\, s_ 3 - 8\, s_ 2^2\, s_
    4)) \, , \cr
		\Delta &= - (c_ 0\, s_ 1^2 - s_ 4^2)^2 \, s_ 3^4 \,s_ 1^2\,  (s_ 1^2\, s_ 2^4 + 4\, c_ 0\, s_ 3^2 - 4\, s_ 2^2\, s_ 3\, s_ 4) \, .
  \end{aligned}
\end{equation}
The factorization of the discriminant and the form of $f$ and $g$ indicate the
following codimension one singularity types and the corresponding gauge
algebras\footnote{We often write $\{p\}$ as a shorthand for $\{p = 0\}$, the
divisor in $B$ cut out by the polynomial $p$.}:
\begin{align}\label{eq:codim1_singularities}
	\begin{array}{rcl}
		\{t\} := \{c_0\,s_1^2 - s_4^2 \} \, : & I_2 \text{ fiber} & \Rightarrow \mathfrak{su}(2) \, \text{ gauge algebra } \, (\mathfrak{su}(2)_A)\\
		\{s_3\} \, : & I_4 \text{ fiber} & \Rightarrow \, \mathfrak{su}(4) \text{ gauge algebra} \\
		\{s_1\} \, : & I_2 \text{ fiber} & \Rightarrow \, \mathfrak{su}(2)
    \text{ gauge algebra } \, (\mathfrak{su}(2)_B) \,.
	\end{array}
\end{align}
Finally, we also have the residual discriminant $\Delta_{\rm res} = s_ 1^2\,
s_ 2^4 + 4\, c_ 0\, s_ 3^2 - 4\, s_ 2^2\, s_ 3\, s_ 4$, which supports
$I_1$ fibers, but no gauge symmetry. We note that the $I_4$ fiber may, at the
level of the vanishing orders, not give rise to an $\mathfrak{su}(4)$ gauge algebra but
instead contribute an $\mathfrak{sp}(2)$ gauge group from monodromy effects
along the divisor $s_3$ \cite{MR0393039,Bershadsky:1996nh,Katz:2011qp}.

Potential matter sits at codimension two loci where irreducible components of
the discriminant intersect each other or self-intersect, and consequently the
singularity type of the fiber enhances.  Explicit computations reveal the
irreducible codimension two loci, with corresponding vanishing orders of
$f,g$, and $\Delta$, summarized in table \ref{tab:codim2_matter_loci}. The table
also contains the matter representations, the origin of which we will discuss now in more
detail.

\begin{table}
	\begin{align*}
		\begin{array}{c|c|c|c}
			\text{Codimension Two Locus} & \text{ord}(f,g,\Delta) & \text{Fiber Type} & 
        \text{Matter Representation}\\ \hline \rule{0pt}{3ex}
			\{t\} \cap \{s_3\} & (0,0,6) & I_6 & ({\bf 2}, {\bf 4}, {\bf 1}) \\
			\{t\} \cap \{s_1\} = \{s_4\} \cap \{s_1\} & (2,3,6) & I_0^* &
      \frac{1}{2} ({\bf 1}, {\bf 1}, {\bf 2})_h  \oplus \frac{1}{2} ({\bf 3}, {\bf 1}, {\bf 2})\\
			\{t\} \cap \{\Delta_{\rm res}\} & (1,2,3) & III & -\\
			\{s_3\} \cap \{s_1\} & (2,3,7) & I_1^* & \frac{1}{2}({\bf 1}, {\bf 6}, {\bf 2})\\
			\{s_3\} \cap \{s_2\} & (2,3,6) & I_0^* & ({\bf 1}, {\bf 6}, {\bf 1}) \\
			\{s_1\} \cap \{c_0\,s_3 - s_2^2\,s_4\} & (0,0,3) & I_3 & ({\bf 1}, {\bf 1}, {\bf 2})
		\end{array}
	\end{align*}
  \caption{Singularity enhancements in codimension two. The corresponding
    matter representations, of the $\mathfrak{su}(2)_A \oplus
    \mathfrak{su}(4)\oplus \mathfrak{su}(2)_B$ gauge algebra, of the 6D F-theory
    compactification are included.
  The $1/2$ in the final column indicates that at each codimension two point
of that kind there exists a half-hypermultiplet instead of a full
hypermultiplet.}\label{tab:codim2_matter_loci} 
\end{table}

From the types of singularity enhancement in table
\ref{tab:codim2_matter_loci}, one can deduce most of the matter
representations right away.  In particular, the enhancements $(I_2, I_4)
\rightarrow I_6$, $I_4 \rightarrow I_0^*$, and $I_2 \rightarrow I_3$ for the
loci $\{t\} \cap \{s_3\}$, $\{s_3\} \cap \{s_2\}$, and $\{s_1\} \cap \{c_0\,s_3
- s_2^2\,s_4\}$ respectively are standard indicators of bifundamental, anti-symmetric
and fundamental matter.  Also, the enhancement $I_2 \rightarrow III$ is
well-known to not support any localized matter.  This explains the
representations $({\bf 2}, {\bf 4}, {\bf 1})$, $({\bf 1}, {\bf 6}, {\bf 1})$
and $({\bf 1}, {\bf 1}, {\bf 2})$ in table \ref{tab:codim2_matter_loci}.

More exotic are the enhancements $(I_4, I_2) \rightarrow I_1^*$ over $\{s_3\}
\cap \{s_1\}$, and especially $(I_2, I_2) \rightarrow I_0^*$ at $\{s_4\} \cap
\{s_1\}$.  Note that the latter locus is also the ordinary double point
singularity of the $\mathfrak{su}(2)_A$ divisor $\{t\}$.  To infer more
information about the representations without resolution, we study the
branching rule for the adjoint representation of the gauge algebra associated
with the enhanced singularity type into the product algebra of the colliding
codimension one divisors.  Hence, at the intersection locus $\{s_3\} \cap
\{s_1\}$ of
the $\mathfrak{su}(4)$ and $\mathfrak{su}(2)_B$ divisors we locally have an
$\mathfrak{so}(10)$, and thus we expect matter in
the $({\bf 1}, {\bf 6}, {\bf 2})$ representation.

At the $I_0^*$ enhancement over $\{s_4\} \cap \{s_1\}$, the local algebra, by
observing the vanishing orders, is
$\mathfrak{so}(8)$.  Since this is an ordinary double point of the
$\mathfrak{su}(2)_A$ divisor $\{t\}$, which is also transversely intersected
by the $\mathfrak{su}(2)_B$ divisor $\{s_1\}$, we have locally the inclusion
$\mathfrak{su}(2)_A \oplus \mathfrak{su}(2)_A \oplus \mathfrak{su}(2)_B
\subset \mathfrak{so}(8)$.  
It is well-known \cite{Sadov:1996zm,Morrison:2011mb} that an $\mathfrak{su}(n)$
self-intersecting in an ordinary double point gives rise to the symmetric and
antisymmetric representations\footnote{As explained in \cite{Morrison:2011mb},
depending on the global structure of the self-intersecting divisor, the
ordinary double point may
instead give rise to the trivial and adjoint representations. This distinction
is not relevant for $\mathfrak{su}(2)$.} of the $\mathfrak{su}(n)$. Since there is an
additional transverse $\mathfrak{su}(2)$ algebra intersecting the
self-intersection point the total matter representation here\footnote{A
  similar situation arose in
  \cite{Klevers:2016jsz}. There, it was a single $\mathfrak{su}(2)$ divisor
  which had a triple self-intersection.  The conclusion is
  that one expects the trifundamental under the local
  $\mathfrak{su}(2)^{\oplus
  3}$ algebra.  In \cite{Klevers:2016jsz}, since all three local
  $\mathfrak{su}(2)$ copies were identified globally, the trifundamental
  decomposes into ${\bf 2} \otimes {\bf 2} \otimes {\bf 2} = {\bf 2} \oplus
  {\bf
  2} \oplus {\bf 4}$.} 
expected is
\begin{equation}
  {\bf 2}_A \otimes {\bf 2}_A
  \otimes {\bf 2}_B = ({\bf 1}_A \oplus {\bf 3}_A) \otimes {\bf 2}_B \cong
  ({\bf
  1}, {\bf 2}) \oplus ({\bf 3}, {\bf 2}) \,,
\end{equation}
where the symmetric and anti-symmetric representations of $\mathfrak{su}(2)_A$
are just the adjoint and trivial representations, respectively.

In order to determine the multiplicities of the matter representations, we
will impose the cancellation of all 6D gauge anomalies. It turns out that this
uniquely fixes all multiplicities to be those shown in table
\ref{tbl:genericmatter}. For more details of the anomaly cancellations in 6D
F-theory compactifications see e.g.~\cite{Kumar:2010ru,Grassi:2011hq}.

If we let $x_{{\bf R}_{ij}}$ denote the number of matter hypermultiplets in a
representation ${\bf R}$ localized along the codimension two points at the
intersection of the divisors $D_i$ and $D_j$ then we can make
the ansatz 
\begin{equation}
  x_{{\bf R}_{ij}} = n_{{\bf R}_{ij}}  D_i \cdot D_j \,.
\end{equation}
In such an ansatz we are careful to distinguish the same representations, for
example the two different ${\bf (1,1,2)}$, that arise from distinct
pairs of intersecting divisors, and may have different coefficients $n_{{\bf
R}_{ij}}$ from each codimension two locus.  Furthermore, we have non-localized
adjoint matter arising as deformation moduli of the gauge algebra divisors.
These are counted by the geometric genus $p_g$ of the divisor
\cite{Witten:1996qb}.  For a smooth divisor $D$, the geometric genus agrees
with the arithmetic genus\footnote{This formula holds of course only for
divisors, i.e., curves, on a twofold base $B$.}
\begin{equation}\label{intersecFormulaDiv}
  p_a = 1 + \frac{1}{2} [D] \cdot ([D] + K_B) \,.
\end{equation}
If $D$ has singularities at points $P_k \in D$, then the two
genera differ by the delta-invariants associated with the
singularities: 
\begin{equation}
  p_g = p_a - \sum_k \delta_k \,.
\end{equation}
For an ordinary double point singularity, which is precisely the singularity
of the $\mathfrak{su}(2)_A$ divisor we are considering, the delta-invariant is
$\delta_k = 1$. The exact multiplicity of the non-localized adjoint matter being the
geometric genus follows from anomaly cancellation, and thus there are no $n_{{\bf
R}_{ij}}$ parameters for this matter. 

Inserting the multiplicities into the anomaly cancellation conditions for all
three gauge factors, it turns out that the anomalies are canceled (independently
of the choices for $\beta, \Sigma$) if and only if the localized matter
multiplicities have
\begin{equation}
  \begin{aligned}
    n_{\bf (2,4,1)} &= 1 &  n_{\bf (1,6,1)} &= 1 & n_{\bf (1,1,2)} &= 1 \cr 
    n_{{\bf (1,1,2)}_h} &= \frac{1}{2} & n_{\bf (3,1,2)} &= \frac{1}{2} &
    n_{\bf (1,6,2)} &= \frac{1}{2} \,,
  \end{aligned}
\end{equation}
where we have used the subscript $h$ to distinguish the two different origins
of ${\bf (1,1,2)}$ matter, consistent with table \ref{tab:codim2_matter_loci}.
The fact that the coefficients $n_{{ \bf R}_{ij}}$ are $1/2$ in some instances
indicates that these are half-hypermultiplets that are situated at those
particular codimension two points. As can be readily observed the
representations associated to the half-hypermultiplets are all pseudo-real,
and thus the half-hypermultiplet exists as a consistent state.
The multiplicities of the matter are summarized in table
\ref{tbl:genericmatter}. 

\begin{table}
  \centering
  \begin{tabular}{c|c|c}
    Fiber Type & Matter & Multiplicity \cr\hline
    $I_2$ & ${\bf (3, 1, 1)}$ & $1 + (\beta + \Sigma) \cdot (2 \beta + K_B + \Sigma)$ \cr 
    $I_4$ & ${\bf (1, 15, 1)}$ & $1 + \frac{1}{2}(\beta + 2 K_B + \Sigma) \cdot (\beta + K_B - \Sigma)$ \cr 
    $I_2$ & ${\bf (1, 1, 3)}$ & $1 + \frac{1}{2}\Sigma \cdot (\Sigma -\beta)$ \cr
    $I_6$ & ${\bf (2, 4, 1)}$ & $2 (\beta + \Sigma) \cdot (-2 K_B - \beta - \Sigma)$ \cr
    $I_0^*$ & ${\bf (1, 1, 2)}_h \oplus {\bf (3, 1, 2)}$ & $\frac{1}{2}\Sigma \cdot(\Sigma + \beta)$ \cr
    $III$ & --- & --- \cr
    $I_1^*$ & ${\bf (1, 6, 2)}$ & $\frac{1}{2} \Sigma \cdot (-2K_B + \beta - \Sigma)$ \cr
    $I_0^*$ & ${\bf (1, 6, 1)}$ & $(- K_B - \Sigma) \cdot (-2K_B - \beta - \Sigma)$ \cr
    $I_3$ & ${\bf (1,1,2)}$ & $\Sigma \cdot (-2K_B + \beta - \Sigma)$
  \end{tabular}
  \caption{The matter multiplicities in the F-theory compactification to 6D on
    the Calabi--Yau elliptic fibration
    (\ref{eq:Weierstrass_functions_generic_solution}). The first three rows
  correspond to the adjoint hypermultiplets arising as deformation modes of
the codimension one components of the discriminant; the remaining rows are
localized codimension two matter.}\label{tbl:genericmatter}
\end{table}

Up until now, we have not discussed how the presence of the $\bbZ_2$ torsional
section affects the F-theory physics.  A perhaps naive expectation, based on
models in previous works \cite{Aspinwall:1998xj} and \cite{Mayrhofer:2014opa},
is that all non-abelian gauge factors should be affected by the $\bbZ_2$
section, i.e., the global structure should be $[SU(2) \times SU(4) \times
SU(2)]/\bbZ_2$. This would be consistent with
the fact that -- other than adjoints -- only either bifundamentals of
$\mathfrak{su}(2)_A \oplus \mathfrak{su}(4)$, or matter carrying the
anti-symmetric ($\bf 6$) representation of $\mathfrak{su}(4)$ exist.  However,
the quotient structure should forbid $\bf (1,1,2)$ matter states, i.e., pure
fundamentals under $\mathfrak{su}(2)_B$.  Given that we have these states, we
propose that the global gauge group is
\begin{align}
  G = \frac{SU(2)_A \times SU(4)}{\bbZ_2} \times SU(2)_B \, .
\end{align}
In addition to the spectrum in table
\ref{tbl:genericmatter}, this observation is also supported by the fact that the torsional section passes through the fiber singularities over the $\mathfrak{su}(2)_A$ and $\mathfrak{su}(4)$ divisor, but no through the $\mathfrak{su}(2)_B$ singularity.
Indeed, the section $S_Q$ \eqref{eqn:WSsecs}, which after solving the $\bbZ_2$ torsional condition $y_Q =0$ \eqref{eq:tuning_generic_solution} sits at
\begin{align}
	[x : y : z] = \left[ \frac{s_1^2\,s_2^2 - 2\,s_3\,s_4}{3} : 0 : 1 \right] \, ,
\end{align}
coincides with the $I_2$ singularity of $\mathfrak{su}(2)_A$ in the Weierstrass model \eqref{eq:Weierstrass_functions_generic_solution} at
\begin{align}
	[x:y:z] = \left[ \frac{s_1^2\,s_2^2 - 2\,s_3\,s_4}{3} : 0 : 1 \right] \quad \text{over} \quad \{c_0 \, s_1^2 - s_4^2\}\, ,
\end{align}
and with the $I_4$ singularity of the $\mathfrak{su}(4)$ divisor at
\begin{align}
	[x:y:z] = \left[ \frac{s_1^2\,s_2^2}{3} : 0 : 1 \right] \quad \text{over} \quad \{s_3\} \, .
\end{align}
However it does not pass through the $I_2$ singularity of $\mathfrak{su}(2)_B$ at
\begin{align}
	[x:y:z] = \left[ \frac{s_3\,s_4}{3} : 0 : 1 \right] \quad \text{over} \quad \{s_1\} \, .
\end{align}
To explicitly verify the global gauge group
structure through homology relation as in
\cite{Mayrhofer:2014opa} would require a global resolution of the model.  While we will not attempt a resolution of the full
model, we will present a resolution for a specialized case in section
\ref{sec:resolution_restricted_model} that exhibits a global gauge group
$[SU(2)_A / \bbZ_2] \times SU(2)_B$.
More precisely, we will consider the resolution of a specialization of
\eqref{eq:Weierstrass_functions_generic_solution}, corresponding to
\begin{equation}\label{eq:simplifiedModel}
  s_2 = 0 \,, \quad s_3 = 1 \,.
\end{equation}
The spectrum of that model will then be determined explicitly, and summarized
in table \ref{tab:spectrum_restricted_model}. It can be easily seen to match
the result in table \ref{tbl:genericmatter} upon imposition of the condition
\begin{equation}\label{eq:condSigmaRestrictedModel}
  \Sigma = -2K_B - \beta \,,
\end{equation}
required for setting $s_3$ to be a constant function.

\subsection[Higgsing the \texorpdfstring{$\bbZ_2$}{Z2} Torsional Model to \texorpdfstring{$U(1)$}{U(1)}]{Higgsing the \boldmath{$\bbZ_2$} Torsional Model to \boldmath{$U(1)$}}\label{sec:matching_spectrum}

One can subsequently use the spectrum of the $\mathbb{Z}_2$ torsional model as
given in table \ref{tab:codim2_matter_loci} to break the group $G=[SU(2)
\times SU(4)]/\mathbb{Z}_2 \times SU(2)$ back to a $U(1)$ with only charge $1$
and $2$ hypermultiplets -- up to an overall normalization -- and verify that
the multiplicities match those found for the singlets in the
Morrison--Park model \eqref{multMP}. 

Using an adjoint field, one can break $G$ to its Cartan subgroup, $U(1)^5$.
There is then a large set of possible ways to break four out of the five
$U(1)$s and obtain a spectrum with only the desired charged hypermultiplets.
One must give vacuum expectation values to four of the many remaining fields,
in such a way as to leave behind only a single $U(1)$ gauge factor and no
remnant discrete symmetries after the Higgsing. An effective approach is to
make use of the Smith normal form \cite{Petersen:2009ip} to keep track of
these subtleties, as well as being implementable algorithmically. We
exhaustively scanned through all of the possibilities, similar in spirit to
the analysis performed in \cite{Baume:2015wia}, and we find that all the
Higgsing chains leading to such a spectrum fall into three distinct classes,
associated to distinct twisting line bundles, $\mathcal{O}(\beta)$, that
characterize the $U(1)$ model.

These three distinct classes of models can be interpreted as different ways
of combining the coefficients, $s_i$, in \eqref{eq:tuning_generic_solution} to
produce, after Higgsing, a Weierstrass model of the form (\ref{eqn:U1WS}).
Higgsing here means that after the $s_i$ are combined into a coefficient
$a(s_i)$ we perform a complex structure deformation which renders $a(s_i)$
generic.

It is evident that we can collect the $s_i$ into $b$, $c_i$, by utilizing the
solution (\ref{eq:tuning_generic_solution}) with which the Morrison--Park
Weierstrass model was tuned to have $\mathbb{Z}_2$ Mordell--Weil torsion in
the first place. The tuned model is characterized by two bundles
$\mathcal{O}(\beta)$ and $\mathcal{O}(\Sigma)$; the twisting line bundle that
characterizes the $U(1)$ model after Higgsing is a tensor product of copies of these bundles,
and the canonical bundle. For the Higgsed $U(1)$ model corresponding to the
identification in (\ref{eq:tuning_generic_solution}) the twisting line bundle
is just $\mathcal{O}(\beta)$.

There are two further collections of coefficients in \eqref{eq:Weierstrass_functions_generic_solution} which give rise to a
Morrison--Park model \eqref{eqn:U1WS}; these are either
\begin{equation}\label{eqn:zz1}
  \begin{aligned}
    b^\prime = s_3 \,, \quad c_0^\prime = s_1^2 c_0 \,,\quad c_1^\prime = 2
    s_1 s_2 s_4 \,,\quad c_2^\prime = s_3 s_4 + s_1^2 s_2^2 \,,\quad
    c_3^\prime = s_1 s_2 s_3 \,,
  \end{aligned}
\end{equation}
or
\begin{equation}\label{eqn:zz2}
  \begin{aligned}
    b^{\prime\prime} =s_1 \,,\quad c_0^{\prime\prime} =c_0 s_3^2 \,,\quad
    c_1^{\prime\prime} =2s_2s_3s_4 \,,\quad c_2^{\prime\prime}
    =s_3s_4+s_1^2s_2^2 \,,\quad c_3^{\prime\prime} =s_1^2s_2 \,,
  \end{aligned}
\end{equation}
with twisting line bundles $\mathcal{O}(\Sigma + \beta)$ and $\mathcal{O}(-2
K_B - \Sigma)$, respectively.
If one was to begin with a $U(1)$ model and enhance according to
(\ref{eqn:zz1}) and (\ref{eqn:zz2}) then these specialization of the
Morrison--Park coefficients are captured in the general solution for
$\mathbb{Z}_2$ Mordell--Weil torsion in (\ref{eq:tuning_generic_solution}).

\section{Resolution of Restricted Model and the Torsional Section}\label{sec:resolution_restricted_model}

In the previous section the structure of the non-simply-connected gauge group
in the effective physics was inferred by observing that the consistency of the
matter spectrum would require that only the $\mathfrak{su}(2)_A$ and the
$\mathfrak{su}(4)$ could be quotiented by $\mathbb{Z}_2$. One can determine
explicitly the action of the $\mathbb{Z}_2$ by studying the crepant
resolution of singularities of the Weierstrass model
(\ref{eq:Weierstrass_functions_generic_solution}). Crepant
resolutions in F-theory
\cite{Esole:2011sm,MS,Krause:2011xj,Lawrie:2012gg,Braun:2014kla,Braun:2015hkv}
allow one to observe physical features which are hidden in the singularities
of the Weierstrass model; in this case we will see explicitly the torsional
relation in homology that is induced by the torsional section.

In the following, we will present the resolution of the restricted version
\eqref{eq:simplifiedModel} of the generic model
\eqref{eq:Weierstrass_functions_generic_solution}. Setting $s_1=0,\,s_3=1$ is
equivalent to setting $c_1=0$ and $c_3 = 0$ in the $U(1)$-fibration
\eqref{eq:MP-fibration}, which is birational to the $U(1)$ Weierstrass model
(\ref{eqn:U1WS}).  Although being specialized, this solution -- leading
to an $\mathfrak{su}(2)_A \oplus \mathfrak{su}(2)_B$ gauge algebra -- exhibits
the two peculiar features of the full model, namely the presence of the
singular point of the $\mathfrak{su}(2)_A$ divisor and the fact that the other
part of the non-abelian gauge group, $\mathfrak{su}(2)_B$, is not affected by the torsional section.
As we will see, these features can be directly extracted from the resolved
fiber structure.

\subsection{Toric Resolution}

\begin{figure}
	\centering
	  \begin{tikzpicture}[scale=1.5]
  \filldraw [ultra thick, draw=black, fill=lightgray!30!white]
      (-1,-1)--(1,0)--(0,1)--(-1,1)--cycle;
    \foreach \x in {-1,0,1}{
      \foreach \y in {-1,0,...,1}{
        \node[draw,circle,inner sep=1.3pt,fill] at (\x,\y) {};
      }
    }

  \draw[ultra thick, -latex]
       (0,0) -- (0,1) node[above right] {$s$};
  \draw[ultra thick, -latex]
       (0,0) -- (-1,1) node[above left] {$u$};
  \draw[ultra thick, -latex]
       (0,0) -- (-1,-1) node[below left] {$v$};
         \draw[ultra thick, -latex]
       (0,0) -- (1,0) node[below right] {$w$};
\begin{scope}[xshift=0.33\textwidth]
\filldraw [ultra thick, draw=black, fill=blue!30!white]
      (-1,-1)--(-1,2)--(0,1)--(1,0)--(0,-1)--cycle;
    \foreach \x in {-1,...,1}{
      \foreach \y in {-1,0,...,2}{
        \node[draw,circle,inner sep=1.3pt,fill] at (\x,\y) {};
      }
    }
	\node [below] at (-1,-1) {$c_3\,u\,v^3$};
	\node [below] at (0,-1) {$b\,v^2\,w$};
	\node [below left] at (-1,0) {$c_2\,s\,u^2\,v^2$};
	\node [below left] at (-1,1) {$c_1\,s^2\,u^3\,v$};
	\node [below left] at (-1,2) {$c_0\,s^3\,u^4$};
	\node [above right] at (0,1) {\textcolor{red}{$s^2\,u^2\,w$}};
	\node [below right] at (1,0) {$s\,w^2$};
	\node [below] at (0,0) {\textcolor{red}{$s\,u\,v\,w$}};
\end{scope}
  \end{tikzpicture}
	\caption{\textit{On the left}: The toric polygon, referred to as $F_6$ in
  \cite{Klevers:2014bqa}, of the fiber ambient space ${\rm Bl}_1 \bbP_{112}$
  of the blown-up $U(1)$ model \eqref{eq:MP-fibration}. \textit{On the right}: The dual polygon, giving rise to the resolved hypersurface equation \eqref{eq:MP_hypersurface_s_resolved}. As pointed out in \cite{Morrison:2012ei}, by assuming a unit coefficient in front of the term $s\,w^2$, the two red monomials can be absorbed by a shift of $w$ by a multiple of $u$.}
	\label{fig:polygon_F6}
\end{figure}
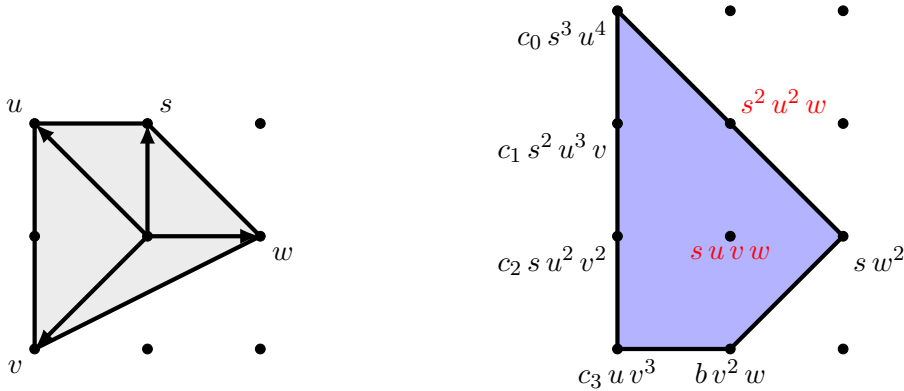

It was shown in the appendix of \cite{Morrison:2012ei} that the codimension
two singularities in the $U(1)$ model \eqref{eq:MP-fibration} can be resolved
torically.  The introduced blow-up divisor, denoted $s$, vanishes
precisely at the rational section in that model. We can write such a model as
a hypersurface $Y$ in a ${\rm Bl}_1 \mathbb{P}_{112}$ fibration $X$ over the
base $B$, given by the equation
\begin{align}\label{eq:MP_hypersurface_s_resolved}
	s \, w^2 + b\,v^2 \,w = c_0\,s^3\,u^4 + c_1\,s^2\,u^3\,v + c_2\,s\,u^2\,v^2 + c_3\,u\,v^3 \, .
\end{align}
This involves blowing up the $\mathbb{P}_{112}$ ambient space at the point
$[0:1:0]$ and taking the proper transform of the hypersurface
(\ref{eq:MP-fibration}).
The Stanley--Reisner ideal of this ambient space can be easily read off from
the toric polygon, shown in figure \ref{fig:polygon_F6}, and is given by
\begin{equation}\label{eq:SR_ideal_BL1P112}
  \{ v \,s, \, u\,w \} \,.
\end{equation}
The divisor classes corresponding to the two sections generating the $U(1)$
are given by 
\begin{equation}
  U := [u] \,, \quad S := [s] \,.
\end{equation}

Upon imposing $c_1=0$ and $c_3 = 0$, the hypersurface
\eqref{eq:MP_hypersurface_s_resolved} develops a further codimension two singularity at
\begin{equation}
  w = s = b = c_2 = 0 \,.
\end{equation}
Such a singularity of $Y$ can be resolved by further blowing up the ambient space at
$w=s=0$, introducing a coordinate $\gamma$, corresponding to a small resolution of the Calabi--Yau hypersurface. After
the blow-up the elliptic fibration $\hat{Y}$ is given by the hypersurface equation
\begin{equation}\label{eq:restricted_example_gamma_resolved}
  \hat{P} \equiv \gamma^2\,s\,w^2 + b\,w\,v^2 - c_0\,\gamma^2 \,s^3 \, u^4 - c_2 \, s \, u^2 \, v^2
  = 0\,,
\end{equation}
in the blown-up ambient space $\hat{X}$ with SR-ideal
\begin{equation}\label{eq:restricted_example_gamma_SRideal}
  \{ s\,v, u\,w, \gamma\, v, w \,s, u\,\gamma \} \,. 
\end{equation}
The discriminant of this fibration will be useful later and is given by
\begin{align}\label{eq:restricted_example_discriminant}
	\Delta_{\hat{Y}} = c_0 \, b^2  \, (c_2^2 - c_0 \, b^2)^2 \, ,
\end{align}
where the component $b$ gives rise to the $\mathfrak{su}(2)_B$ algebra and
$(c_2^2 - c_0 \, b^2)$ to the $\mathfrak{su}(2)_A$.

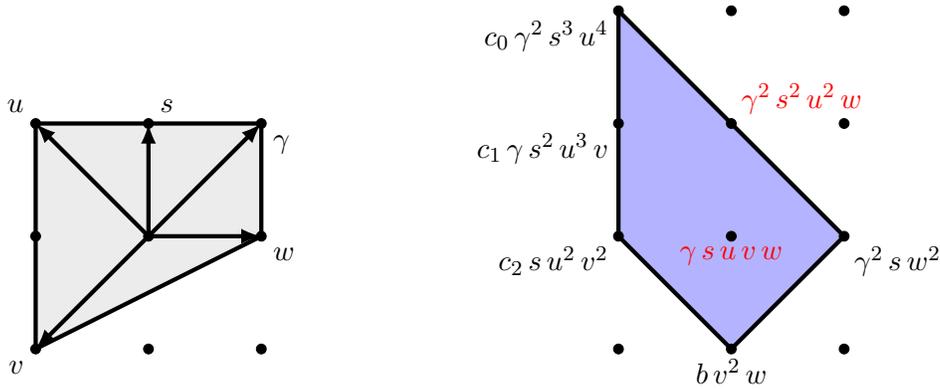
\begin{figure}
	\centering
	  \begin{tikzpicture}[scale=1.5]
  \filldraw [ultra thick, draw=black, fill=lightgray!30!white]
      (-1,-1)--(-1,1)--(1,1)--(1,0)--cycle;
    \foreach \x in {-1,0,1}{
      \foreach \y in {-1,0,...,1}{
        \node[draw,circle,inner sep=1.3pt,fill] at (\x,\y) {};
      }
    }

  \draw[ultra thick, -latex]
       (0,0) -- (0,1) node[above right] {$s$};
  \draw[ultra thick, -latex]
       (0,0) -- (-1,1) node[above left] {$u$};
  \draw[ultra thick, -latex]
       (0,0) -- (-1,-1) node[below left] {$v$};
  \draw[ultra thick, -latex]
       (0,0) -- (1,0) node[below right] {$w$};
  \draw[ultra thick, -latex]
       (0,0) -- (1,1) node[below right] {$\gamma$};
\begin{scope}[xshift=0.33\textwidth]
\filldraw [ultra thick, draw=black, fill=blue!30!white]
      (-1,0)--(-1,2)--(0,1)--(1,0)--(0,-1)--cycle;
    \foreach \x in {-1,...,1}{
      \foreach \y in {-1,0,...,2}{
        \node[draw,circle,inner sep=1.3pt,fill] at (\x,\y) {};
      }
    }
	\node [below] at (0,-1) {$b\,v^2\,w$};
	\node [below left] at (-1,0) {$c_2\,s\,u^2\,v^2$};
	\node [below left] at (-1,1) {$c_1\,\gamma\,s^2\,u^3\,v$};
	\node [below left] at (-1,2) {$c_0\,\gamma^2\,s^3\,u^4$};
	\node [above right] at (0,1) {\textcolor{red}{$\gamma^2\,s^2\,u^2\,w$}};
	\node [below right] at (1,0) {$\gamma^2\,s\,w^2$};
	\node [below] at (0,0) {\textcolor{red}{$\gamma\,s\,u\,v\,w$}};
\end{scope}
  \end{tikzpicture}
	\caption{\textit{On the left}: The toric polygon, called $F_8$ in
  \cite{Klevers:2014bqa}, of the $\gamma$-blow-up of ${\rm Bl}_1 \bbP_{112}$.
  \textit{On the right}: The dual polygon, giving rise to the hypersurface
  equation for $\hat{Y}_{F_8}$. The $\gamma$-blow-up removes the vertex of the dual polygon
  corresponding to the $c_3$-term of the $U(1)$ model
  \eqref{eq:MP_hypersurface_s_resolved}. As discussed in
  \cite{Klevers:2014bqa}, this elliptic fibration has an $I_2$ locus above
$b=0$. Were it not for a unit coefficient in front of $\gamma^2\,s\,w^2$
(which allows us to absorb the red terms), there would be another $I_2$ locus present.
	The non-toric tuning $c_1 \rightarrow 0$ enhances the $U(1)$ ``torsionally''.}
	\label{fig:polygon_F8}
\end{figure}

The blow-up $\gamma$ can be also engineered torically. 
In terms of the toric diagram of the fiber ambient space, this blow-up
precisely corresponds to introducing an additional ray between the rays of $w$
and $s$, as one can see in figure \ref{fig:polygon_F8}. This removes a vertex of the dual
polygon, effectively setting $c_3=0$ in \eqref{eq:MP_hypersurface_s_resolved} and defining a new hypersurface $\hat{Y}_{F_8}$.  The blow-up $\gamma$ defines a section
$\Lambda = [\gamma]$ which generates a $U(1)$ in the F-theory compactification on $\hat{Y}_{F_8}$.  Thus, we can
understand $\hat{Y}$ as a non-toric restriction of the generic toric
hypersurface\footnote{That is, generic up to the constant coefficient of the
$\gamma^2 \, s \, w^2$ term.} $\hat{Y}_{F_8}$ by $c_1 \rightarrow 0$.  It can be easily shown that this tunes the
section $\Lambda$ to be $\bbZ_2$ torsional, thus enhancing the $U(1)$ to a
non-abelian symmetry.

\subsection{Torsional Section and Global Gauge Group Structure}

With the fully resolved elliptic fibration \eqref{eq:restricted_example_gamma_resolved}, we want to explicitly determine the homological relation that leads to the non-trivial global gauge group structure.
First, it is straightforward to verify the $\mathfrak{su}(2)_B$ symmetry localized over $b=0$.
Over this locus, the resolved hypersurface \eqref{eq:restricted_example_gamma_resolved} factorizes as
\begin{align}\label{eqn:mildred}
	\hat{P}|_{b=0} = s \left(\gamma^2\, w^2-u^2 \left(c_0 \,\gamma^2\,\,s^2 \, u^2 +c_2 \, v^2 \right)\right) \, ,
\end{align}
thus showing that the section $S$ which generated the $U(1)$ in the Morrison--Park model \eqref{eq:MP_hypersurface_s_resolved}, becomes an exceptional divisor.
It is a ruled surface where the $\mathbb{P}^1$ fiber has positive intersection
with the zero-section, $U$, and so we refer to this fibral curve as the affine
node of the $I_2$ fiber.
The Cartan generator of $\mathfrak{su}(2)_B$ is given by the divisor
corresponding to the second component of (\ref{eqn:mildred}), with class
\begin{equation}
  E_B = [b] - S \,,
\end{equation}
which is by definition the remainder of the total divisor class of
(\ref{eqn:mildred}) after the exceptional divisor corresponding to the affine
node has been subtracted off.

The fiber splitting over the $\mathfrak{su}(2)_A$ locus can be described through prime ideals.
Specifically, one of the two prime factors of the ideal $(\hat{P}, \, b^2\,c_0 - c_2^2)$ is generated by four polynomials,
\begin{align}
	\begin{split}
		I = \left( b^2\,c_0 - c_2^2 \, , \quad c_2 \,s \, u^2 - b\,w \, , \quad b\,c_0\,s\,u^2 - c_2\,w \, , \quad c_0 \, s^2 \,u^4 - w^2 \right) \, .
	\end{split}
\end{align}
This codimension two subvariety $V(I)$ of the ambient space $\hat{X}$ is a divisor of $\hat{Y}$ localized over $\{b^2\,c_0 - c_2^2\}$. 
The intersection of this subvariety with $u=0$ would require $w=0$
as well, however, since $u\,w$ is in the SR-ideal
\eqref{eq:restricted_example_gamma_SRideal} this intersection is empty.  In
other words, the zero-section, $U$, does not intersect $V(I)$, which hence
corresponds to the non-affine Cartan divisor of $\mathfrak{su}(2)_A$.  Its homology class
can be extracted using prime ideal techniques (see, for example, the appendix of
\cite{Lin:2016vus}), yielding
\begin{align}
	\begin{split}
		E_A = [V(I)] = ([b] + [w]) \cdot_{\hat{X}} ([c_2] + [w]) - [b] \cdot_{\hat{X}}
    [c_2] = - 4\,(2\,U + S + \beta) \cdot_{\hat{X}} K_B \, ,
	\end{split}
\end{align}
where $\cdot_{\hat{X}}$ denotes the intersection product in $\hat{X}$, and
$K_B$ now abusively denotes the pullback of the canonical class of $B$ to $\hat{X}$.
In terms of the ambient space homology, one can now use the linear equivalence and SR-ideal relations to show that
\begin{align}
	[\hat{P}] \cdot_{\hat{X}} (\Lambda - U + K_B) + \frac{1}{2} [V(I)] = 0 \, .
\end{align}
By another abuse of notation, we will use the same label for (toric) divisors of the ambient space $\hat{X}$ and their pull-backs to the hypersurface.
Then, the above equation implies that, in the homology of $\hat{Y}$, we have
\begin{align}\label{eq:torsional_section_homology_relation}
	\Lambda - U + K_B = - \frac{1}{2} E_A\, .
\end{align}
This relation is the origin of the non-trivial global structure of the $\mathfrak{su}(2)_A$ factor \cite{Mayrhofer:2014opa}, which we will review briefly for the case at hand.
Suppose we have matter states $\bf w$ from M2-branes wrapping a fibral curve $C$ in the elliptic fibration $\hat{Y}$.
Because of the relation \eqref{eq:torsional_section_homology_relation}, the Cartan charge $q = E_A \cdot C$ of the state $\bf w$ under $\mathfrak{su}(2)_A$ satisfies
\begin{align}
	-\frac{q}{2} = C \cdot (\Lambda - U + K_B) \, ,
\end{align}
where $\cdot$ now denotes the intersection product on $\hat{Y}$.  Since $C$ is
a fibral curve (i.e., localized over a point in the base), the intersection
with the pullback of a base divisor, like $K_B$, vanishes.  We are left with the conclusion that
$C\cdot (\Lambda - U) = - \frac{q}{2}$.  Now $C$, $\Lambda$ and $U$ -- being
classes of subvarieties of $\hat{Y}$ -- are integral in homology.  And since
$\hat{Y}$ is smooth by construction, its intersection pairing must be
integral, forcing $q/2$ to be an integer.  This implies that we cannot have
any representations with odd charges under the $\mathfrak{su}(2)_A$ Cartan
generator.  In other words, the global structure of this gauge factor is
$SU(2)/\bbZ_2 \cong SO(3)$.  The torsional homology relation
\eqref{eq:torsional_section_homology_relation} does not involve the
$\mathfrak{su}(2)_B$ divisor, so we have no such a restriction on the
allowed representations of $\mathfrak{su}(2)_B$.  Hence, we find that the
geometry of the resolved elliptic fibration
\eqref{eq:restricted_example_gamma_resolved} explicitly accounts for the
global gauge group structure $SU(2)/\bbZ_2 \times SU(2)$.

\subsection{Matter States and Codimension Two Enhancements}\label{sec:codim_2_resolved_res_example}

We proceed with analyzing the matter enhancements and confirm the previous
results (cf.~table \ref{tbl:genericmatter} with the condition
\eqref{eq:condSigmaRestrictedModel}) found through anomaly cancellation.  From
the discriminant \eqref{eq:restricted_example_discriminant}, we immediately
see that the singularity enhances over the codimension two loci:
\begin{align}
	c_0 = c_2 = 0 \, , \quad b=c_0=0 \, , \quad b=c_2 = 0 \, .
\end{align}
Over the first locus resides a type $III$ fiber, consisting of two fiber $\bbP^1$
components intersecting each other in a double point.  In F-theory, a
codimension two enhancement from $I_2$ to type $III$ hosts no matter.

The second locus, $b=c_0=0$, lies on the $\mathfrak{su}(2)_B$ divisor $b=0$.
Here, we find an enhancement to $I_3$, and thus we expect fundamentals of
$\mathfrak{su}(2)_B$ that are uncharged under $\mathfrak{su}(2)_A$.
Concretely, setting $b$ and $c_0$ to zero in $\hat{P}$ yields
\begin{align}
	\hat{P}|_{b=c_0=0} = -s\, (c_2\,u^2\,v^2 - \gamma^2\,w^2) = -s \, \underbrace{(\sqrt{c_2}\,u\,v + \gamma\,w)}_{\bbP^1_+}\,\underbrace{(\sqrt{c_2}\,u\,v - \gamma\,w)}_{\bbP^1_-} \, .
\end{align}
Note that the factorization of the quadratic term into $\bbP^1_{\pm}$ involves
taking the square root of $c_2$ in codimension two, which is generic on a
twofold base.  It is then straightforward to compute the intersection numbers
with the divisor $E_B$ associated with the Cartan generator of
$\mathfrak{su}(2)_B$.  One readily finds 
\begin{equation}
  E_B \cdot [\bbP^1_{\pm}] = -1 \,,
\end{equation}
that is, states which transforms as weights in the fundamental representation
of $\mathfrak{su}(2)_B$.  Because there are two distinct fiber
components at $b=c_0=0$ having fundamental Cartan charges, that can be wrapped
individually by M2- and anti-M2-branes, the corresponding 6D theory from
F-theory compactified on a threefold $\hat{Y}$ has a full hypermultiplet of
states in the fundamental representation of $\mathfrak{su}(2)_B$ localized at
$b=c_0=0$.

\begin{figure}
  \centering
  \def\svgwidth{0.25\vsize}
  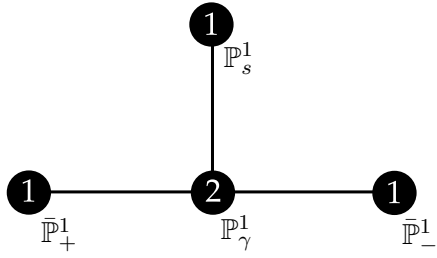
  \caption{The structure of the codimension two singular fiber over the locus
  $b = c_2 = 0$, demonstrating the intersection pattern of the curves in
  (\ref{eq:restricted_example_codim_2_locus_splitting}); the numbers on each
  node indicate the multiplicity. It is a non-Kodaira singular fiber which is a
contraction of the $I_0^*$ Kodaira fiber where one multiplicity one node is
removed.}\label{fig:so8}
\end{figure}

Finally, let us examine the third enhancement locus at $b=c_2=0$.  This locus
lies at the intersection of both $\mathfrak{su}(2)$ divisors, so we expect to
find matter that potentially carry non-trivial representations under both
gauge factors.  The hypersurface equation at that point factors as
\begin{equation}\label{eq:restricted_example_codim_2_locus_splitting}
  \hat{P}|_{b=c_2=0} = s\,\gamma^2 \, (w^2 - c_0 \, s^2 \, u^4) = \underbrace{s}_{\bbP^1_s} \, \underbrace{\gamma^2}_{\bbP^1_\gamma} \, \underbrace{(w + \sqrt{c_0} \, s\,u^2)}_{\bar{\bbP}^1_+} \, \underbrace{(w - \sqrt{c_0} \, s \, u^2)}_{\bar{\bbP}^1_-} \, . 
\end{equation}
Considering the SR-ideal \eqref{eq:restricted_example_gamma_SRideal}, it is
straightforward to show that the components intersect each other in an affine
$\mathfrak{so}(8)$ Dynkin diagram with one external node removed, as can be seen in
figure \ref{fig:so8}. The component $\bbP^1_\gamma$ is the central node with
multiplicity two, and $\bbP^1_s$ is the affine node intersected by the
zero-section. These kinds of non-Kodaira fibers have been observed before and
can be understood by studying the Coulomb branch of the associated M-theory
compactification on the resolved geometry
\cite{Hayashi:2013lra,Hayashi:2014kca,Esole:2014bka,Esole:2014hya}. In
\cite{Hayashi:2014kca}, it was noted that the non-Kodaira singular fibers in
codimension two have the form of contractions of Kodaira fibers (see also
\cite{Cattaneo:2013vda}), and this is consistent with what is observed here.  A
key point is that this particular fiber, where one specific node is deleted is
related to the choice of resolution; topologically distinct crepant
resolutions give rise to contracted $I_0^*$ fibers with different nodes
removed.

In order to compute the Cartan charge of the fiber components under
$\mathfrak{su}(2)_A$, we will use the result
\eqref{eq:torsional_section_homology_relation} of the previous subsection,
\begin{equation}
  E_A = 2\,(U - K_B -\Lambda) \,.
\end{equation}
With that, we can easily compute
intersection numbers in the ambient space homology to obtain
\begin{equation}
	\begin{split}
		(E_A, E_B) \cdot [\bbP^1_\gamma] & = (2,-1) \, , \\
		(E_A, E_B) \cdot [\bar{\bbP}^1_\pm] & = (-2,0) \, .
	\end{split}
\end{equation}
Therefore, omitting the effective curves whose homology class is not localized
in codimension two, the full set of effective genus zero fiber curves 
that are localized at $b=c_2=0$ can be summarized as follows:
\begin{align}\label{tab:restricted_example_localized_fiber_curves}
	\begin{array}{c|c|c}
    \text{Curve} & \text{Cartan Charges} & \text{Representation} \\ \hline\hline \rule{0pt}{3ex}
		\bbP^1_\gamma & (2,-1) & ({\bf 3,2}) \\
		\bbP^1_\gamma + \bar{\bbP}^1_+ & (0,-1) & ({\bf 3,2}) \\
		\bbP^1_\gamma + \bar{\bbP}^1_+ + \bar{\bbP}^1_-& (-2,-1) & ({\bf 3,2}) \\ \hline \rule{0pt}{2.5ex}
		\bbP^1_\gamma + \bar{\bbP}^1_- & (0,-1) & ({\bf 1, 2})
	\end{array}
\end{align}
Note that both combinations $\bbP^1_\gamma + \bar{\bbP}^1_\pm$ have the same
Cartan charges, and it is a matter of choice into which representation we put
each; however, full gauge invariance requires that one is in $(\bf 3,2)$ and
one in $(\bf 1,2)$.\footnote{ From a different point of view, the
  $\mathfrak{su}(2)_A$ Cartan charges can be interpreted as those of states in
  the tensor product ${\bf 2} \otimes {\bf 2} \cong {\bf 3} \oplus {\bf 1}$,
which naturally has two states with zero Cartan charge.} It can be observed
that for each
representation, the number of corresponding effective curves is half the
dimension of the representation.  This reflects the result we found
through anomaly cancellation, see tables \ref{tab:codim2_matter_loci} and
\ref{tbl:genericmatter}, namely that each point in $\{b = c_2 =0\}$ supports
only a half-hypermultiplet in the representations $(\bf 3,2)$ and $(\bf 1,2)$.

The spectrum arising from singular fibers in F-theory compactified on
$\hat{Y}$ is summarized in table \ref{tab:spectrum_restricted_model}. We have
included for later convenience the multiplicities of hypermultiplets, which 
 are just the restriction of those in the generic model
\eqref{eq:Weierstrass_functions_generic_solution} to $\Sigma = -2K_B - \beta$.
Thus, it is also no surprise that the gauge anomalies all cancel.
\begin{table}
	\centering
	\begin{tabular}{c|c|c|c}
	Locus & Fiber Type & Matter Rep. & Multiplicity of Hypermultiplets \\ \hline \rule{0pt}{3ex}
		$\{c_0\}$ & $I_1$ & $-$ & $-$  \\
		$\{c_2^2 - c_0\,b^2\}$ & $I_2$ & ${\bf (3,1)}$ & $1 + 2K_B^2 - 2K_B\cdot \beta$  \\
		$\{ b\}$ & $I_2$ & ${\bf (1,3)}$ & $1+ K_B^2 +\frac{1}{2} (\beta^2 + 3K_B\cdot \beta)$  \\
		$\{c_0\} \cap \{c_2\}$ & $III$ & $-$ & $-$ \\
		$\{c_0\} \cap \{b\}$ & $I_3$ & ${\bf (1,2)}$ & $-4K_B \cdot \beta - 2\beta^2$ \\
		$\{b\} \cap \{c_2\}$ & $I_0^* ({\rm reduced})$ & ${\bf (3 \oplus 1,2)}$ &
    $2K_B^2 + K_B\cdot\beta$			
	\end{tabular}
	\caption{Representations and multiplicities of the matter associated to
    codimension two singularities for the restricted model \eqref{eq:simplifiedModel}.}
	\label{tab:spectrum_restricted_model}
\end{table}

\subsection{Cancellation of Gravitational Anomaly}\label{sec:grav_anomaly_specialized_model}

In section \ref{sec:F-theory_on_Z2}, we have used gauge anomalies to determine
the spectrum.  Here, we will discuss the cancellation of the gravitational
anomaly; in anticipation of the discussion in section \ref{sec:rosie} we shall
include several explicit details.  In contrast to gauge
anomalies, the gravitational anomaly is sensitive to uncharged hypermultiplets.  For
F-theory compactifications on a smooth elliptic Calabi--Yau threefold $\hat{Y}
\rightarrow B$, the number of uncharged hypermultiplets is $1 +
h^{2,1}(\hat{Y})$.  To compute $h^{2,1}(\hat{Y})$, we employ the standard
relation
\begin{align}\label{eq:relation_chi_h_smooth_CY3}
	\chi_{\rm top}(\hat{Y}) = 2 \left( h^{1,1}(\hat{Y}) - h^{2,1}(\hat{Y})\right) \, .
\end{align}
The topological Euler characteristic $\chi_{\rm top}$ can be determined by
dividing $\hat{Y}$ into subspaces and using the additive property of
$\chi_{\rm top}$.  This simplifies drastically if we can choose the subspaces
such that they are all product spaces, in which case the Euler characteristic
just becomes the product of $\chi_{\rm top}$ for the factors.  For the elliptic
fibration $\hat{Y}$, the singular fibers
in table \ref{tab:spectrum_restricted_model} provide a natural
division of $\hat{Y}$ into subspaces \cite{Grassi:2011hq,
Arras:2016evy,Esole:2017kyr}. The dramatic simplification that occurs when
considering an elliptic fibration can be summarized by noting that
\begin{equation}
  \chi_{\rm top}(I_0) = 0 \,,
\end{equation}
that is, the generic fiber, which has a smooth torus, or $I_0$ fiber, has
vanishing Euler characteristic. Due to this the only subspaces that
contribute to the Euler characteristic are those which involve the singular
fibers. Specifically, the decomposition of the Euler characteristic is in
terms of the following two classes of contributions.

\noindent\textbf{\uline{Codimension two}}

Here the subspaces are of the form ${\rm pt} \times {\rm fiber}$, so that 
    \begin{equation}
      \chi_{\rm top}({\rm pt} \times {\rm fiber}) = \chi_{\rm top} ({\rm
      fiber}) \,.
    \end{equation}
    For singular fibers consisting of smooth $\bbP^1$s with normal crossing
    intersections, the Euler characteristic can be computed by adding the
    contributions of each $\bbP^1$, which is 2 minus the number of
    intersection points on that $\bbP^1$, and then add the total number of
    intersection points in that fiber.  For our model, we have type $III$
    fibers with $\chi_{\rm top} = 3$, $I_3$ fibers which have $\chi_{\rm
    top}=3$, and the reduced $I_0^*$ fibers with $\chi_{\rm top}=5$.  Thus,
    the total contribution of codimension two fibers to $\chi_{\rm
    top}(\hat{Y})$ is the number of points in the base with these specific
    fiber types, see table \ref{tab:spectrum_restricted_model}:
	\begin{align}\label{eq:chi_contribution_codim2}
		\chi^{(\rm codim \, 2)} = 3\,[c_0] \cdot [c_2] + 3\,[c_0] \cdot [b] +
    5\,[b] \cdot [c_2] = 20\,K_B^2 - 14\, K_B\cdot \beta - 6\,\beta^2 \, .
	\end{align}
  
\noindent\textbf{\uline{Codimension one}}
  
The codimension one subspaces are ruled surfaces of the form $\Sigma_i \times
{\rm fiber}_i$, where $\Sigma_i$ are the discriminant components with fiber
type $i$. Thus the contribution to the Euler characteristic from these
singular fibers is
    \begin{equation}
      \chi_{\rm top}(\Sigma \times {\rm fiber}) = \chi_{\rm top}(\Sigma) \chi_{\rm top} ({\rm
               fiber}) \,.
    \end{equation}
The topological Euler characteristic of $\Sigma$ is given by 
    \begin{equation}
      \chi_{\rm top}(\Sigma) = -(\Sigma_i + K_B)\cdot \Sigma_i + \sum_s \epsilon_s\cdot \#(P_s) \,.
    \end{equation}
The points $P_s$ are the codimension two enhancement points, of fiber type
$s$, that have already been accounted for, and give rise to the correction
term $\sum_s \epsilon_s\cdot \#(P_s)$.  The value of $\epsilon_s$ depends on
the singularity structure of $\Sigma_i$ at $P_s$ in the base
\cite{Arras:2016evy}.  For the case at hand,
in table \ref{tab:spectrum_restricted_model}, we note that $\epsilon = -1$ for the
enhancement points of type $III$ and $I_3$ fibers on any affected discriminant
component.  On the other hand, the coefficient for the $I_0^*$ enhancement
point depends on the divisor $\Sigma$ for which we are considering the
contribution: For $\Sigma_B = \{b\}$, we have $\epsilon_{I_0^*}=-1$, whereas
for $\Sigma_A = \{c_2^2 - c_0\,b^2\}$ the ordinary double point on the divisor
gives $\epsilon_{I_0^*}=0$.  As a last ingredient, we need that $\chi_{\rm
top} = 1$ for singular $I_1$ fibers.  In summary, we have the following
codimension one contributions:
	\begin{align}\label{eq:chi_contribution_codim1}
		\begin{split}
			\chi^{(\rm codim \,1)}_{c_0} & = -( [c_0] + K_B)\cdot [c_0] - [c_0]
      \cdot ([c_2] + [b]) = 6\,K_B\cdot \beta - 2\,\beta^2 \, , \\
			\chi^{(\rm codim \,1)}_{c_2^2 - c_0\,b^2} & = 2\,(- ([c_2^2] + K_B)
      \cdot [c_2^2] - [c_0] \cdot [c_2]) = -24 \, K_B^2 + 8 \, K_B\cdot \beta \, , \\
			\chi^{(\rm codim \,1)}_{b} & = 2 \, (- ([b] + K_B) \cdot [b] - [b]\cdot
      ([c_0]+[c_2])) = -12 \, K_B^2 - 2\,K_B\cdot \beta + 2 \, \beta^2 \, .
		\end{split}
	\end{align}
Thus, the Euler
characteristic of $\hat{Y}$, which is the sum of
\eqref{eq:chi_contribution_codim2} and \eqref{eq:chi_contribution_codim1}, is
\begin{align}\label{eq:chi_su2xsu2}
	\chi_{\rm top}(\hat{Y}) = - 16\,K_B^2 - 2\, K_B\cdot\beta - 6\,\beta^2\, .
\end{align}

To employ the relationship \eqref{eq:relation_chi_h_smooth_CY3} between
$\chi_{\rm top}(\hat{Y})$ and the Hodge numbers, we now only need to know $h^{1,1}(\hat{Y})$, which
by the Shioda--Tate--Wazir theorem \cite{MR2041769} is
\begin{align}\label{eq:shioda_tate_wazir_with_section}
\begin{split}
	h^{1,1}(\hat{Y}) & = 1 + h^{1,1}(B) + {\rm rk}(G) \\
	& = 13 - K_B^2 \, ,
\end{split}
\end{align}
where we have used that for a twofold base B, $h^{1,1}(B) = 10 - K_B^2$.
This determines the number of uncharged hypermultiplets:
\begin{align}\label{eq:uncharged_hypers_su2xsu2}
	n_H^0 = 1+h^{2,1}(\hat{Y}) = 1+ h^{1,1}(\hat{Y}) - \frac{\chi_{\rm top}}{2}
  = 14 + 7\,K_B^2 + K_B\cdot \beta+3\,\beta^2 \, .
\end{align}
Meanwhile, the number of charged hypermultiplets is given in table
\ref{tab:spectrum_restricted_model}, and  a quick counting
yields\footnote{Note that the uncharged (Cartan) states of the codimension one
  deformation modes are accounted for in the $h^{2,1}(\hat{Y})$ uncharged
  hypermultiplets. Hence, each codimension one adjoint representation of
  $\fkg$ only contributes an additional $\dim (\fkg) - {\rm rk}(\fkg)$
  hypermultiplets to the gravitational anomaly.}
\begin{align}\label{eq:charged_hypers_su2xsu2}
	n_H^c = 4 + 22\, K_B^2 - K_B\cdot \beta-3\,\beta^2 \, .
\end{align}
The final contributions to the anomaly come from the six vector multiplets of
the $\mathfrak{su}(2) \oplus \mathfrak{su}(2)$ gauge fields, and $n_T =
h^{1,1}(B) - 1 = 9 - K_B^2$ tensor multiplets\footnote{Tensor multiplets are
  related to strings in the 6D theory, the worldvolume theories of which have
  recently been studied in
  \cite{Haghighat:2015ega,Lawrie:2016axq,Lawrie:2016rqe,Couzens:2017way}.} from divisors in the base.
Thus, we verify that the gravitational anomaly cancels \cite{Kumar:2010ru}:
\begin{align}
	(n_H^0 + n^c_H) - n_V + 29\,n_T = 18 + 29\,K_B^2 - 6 + 29\,(9-K_B^2) = 273 \, .
\end{align}

\section{Mordell--Weil Torsion in the Presence of Bisections}\label{sec:rosie}

It has been observed in examples
\cite{Klevers:2014bqa,Oehlmann:2016wsb,Cvetic:2016ner} that an elliptic
fibration, $Y$, with Mordell--Weil torsion $\bbZ_n$ is ``dual'' to a
genus-one fibration ${Y}^\vee$ with an $n$-section.
To such a multi-section geometry, one can associate the Tate--Shafarevich group, $\Sh(J({Y}^\vee))$, consisting of the set of all genus-one
fibrations, without isolated multiple fibers, which share the same Jacobian
fibration $J(Y^\vee)$ as $Y^\vee$ \cite{MR0106226,MR0162806,MR1242006}.  For a
genus-one fibration $Y^\vee$ with an independent $n$-section and no
codimension one singularities, 
\begin{align}
	\Sh(Y^\vee) = \bbZ_n
\end{align}
is believed to encode the discrete $\bbZ_n$ symmetry of F-theory compactified
on $J(Y^\vee)$ \cite{Braun:2014oya, Mayrhofer:2014laa, Morrison:2014era,
Cvetic:2015moa}.  
In the presence of codimension one singularities, one observes that depending on the non-abelian gauge algebra, the discrete symmetry can be enhanced by the center \cite{Garcia-Etxebarria:2014qua, Klevers:2014bqa, Oehlmann:2016wsb}.
In any case, the common folklore is that while the
M-theory is different, F-theory compactified on any genus-one fibration in
$\Sh(J(Y^\vee))$ gives the same field theory as $J(Y^\vee)$. 
Thus, the conjecture is that for an F-theory compactification on
$Y$ with non-trivial global gauge group $G/\bbZ_n$, there is a dual
compactification on an $n$-section geometry $Y^\vee$.

So far, the conjecture \cite{Oehlmann:2016wsb} is based on a set of toric
examples \cite{Klevers:2014bqa, Braun:2014qka}. For some of these there is a
dual heterotic description \cite{Cvetic:2016ner}, where this duality can be
understood rigorously.  In these examples, the duality manifests itself as a
``fiberwise mirror symmetry'': $Y$ and $Y^\vee$ are generic complete
intersections in an ambient space $X$ resp.~$X^\vee$, which are fibrations of
a toric fiber ambient space $\cal A$ resp.~${\cal A}^\vee$ that are mirror
to each other (i.e., they have dual toric fans).  However, it is currently not
known how to generalize the duality to non-toric examples, mainly because
there are no known such constructions.

In this section, we provide evidence that a model relevant in this context arises by deforming the
geometry discussed in section \ref{sec:resolution_restricted_model}. In
particular, we propose that our non-toric construction yields a bisection geometry $Y_b$ and an associated Jacobian fibration $J(Y_b)$, whose F-theory compactification has gauge group
\begin{align}
	\frac{SU(2)}{\bbZ_2} \times \bbZ_2 \, .
\end{align}
Having both a bisection and $\bbZ_2$ Mordell--Weil torsion in the Jacobian, the construction may be ``self-dual'' in the above
sense, although we will not explore this direction in the present work.\footnote{Note that such self-dual examples also exist in the list of toric models
\cite{Oehlmann:2016wsb}.}
Nevertheless, a better understanding of this model might shed light on a non-toric formulation of the duality.
However, it turns out that just interpreting F-theory on the pair $(J(Y_b), Y_b)$ is more intricate than expected.
In the following, we will see that such an interpretation will require further conceptual understanding of F-theory compactifications on multi-section geometries and their associated Jacobians.

\subsection{A Jacobian Fibration with Torsional Section}

It is well known \cite{Braun:2014oya, Anderson:2014yva,
Garcia-Etxebarria:2014qua, Mayrhofer:2014haa, Mayrhofer:2014laa} that the
Morrison--Park model can be deformed through a conifold transition, which
physically breaks the $U(1)$ to a $\bbZ_2$ symmetry by giving states with
charge 2 a non-zero vacuum expectation value.  In the Weierstrass form
\eqref{eqn:U1WS}, the deformation $b^2 \rightarrow 4\,c_4$ yields a new
Weierstrass model
\begin{equation}\label{eq:standard_bisection_model_jacobian}
  y^2 = x^3 + \left( c_1 c_3 - \frac{1}{3}c_2^2 - 4 c_0 c_4 \right) x\,z^4 + \left(- c_0c_3^2 +
  \frac{1}{3}c_1c_2c_3 - \frac{2}{27}c_2^3 + \frac{8}{3}c_0c_2c_4 - c_1^2
  c_4 \right) z^6
  \, .
\end{equation}
This elliptic fibration has non-trivial $\bbZ_2$ torsional three-cycles
\cite{Mayrhofer:2014laa}, which indicates the existence of a discrete $\bbZ_2$
symmetry already in the M-theory compactification, and which uplifts to the
F-theory compactification.  Likewise, there are
terminal singularities in codimension two of the fibration, which corresponds
to matter charged only under the $\bbZ_2$ \cite{Braun:2014nva, Braun:2014oya,
Morrison:2016lix, Arras:2016evy}.  This geometry is the Jacobian
$J(Y_{\bbZ_2})$ of a generic hypersurface $Y_{\bbZ_2}$ in a $\bbP_{112}$
fibration:
\begin{align}\label{eq:generic_P112_hypersurface}
	Y_{\bbZ_2} : \qquad w^2 = c_0 \, u^4 + c_1\,u^3\,v + c_2\,u^2\,v^2 + c_3\,u\,v^3 + c_4\,v^4 \, .
\end{align}
One can easily check that by (the toric) tuning $c_4 \rightarrow b^2/4$ (and a subsequent coordinate shift), this hypersurface becomes the $\bbP_{112}$ description \eqref{eq:MP-fibration} of the Morrison--Park model.
Note that this also identifies the classes of $c_i, i=0,...,3$ with those of the Morrison--Park model \eqref{tab:coefficients_classes_MP}, and $[c_4] = 2[b]$.
Unlike the Jacobian geometry, this hypersurface is a smooth genus-one fibration with a bisection and trivial torsional homology.
The apparently missing $\bbZ_2$ symmetry in M-theory compactified on
$Y_{\bbZ_2}$ is restored only when we perform the F-theory uplift, in which
case the discrete symmetry emerges as a subgroup of the Kaluza--Klein $U(1)$.
While it is not instructive to present explicitly the discriminant and the
enhancement loci, we do highlight that F-theory on $J(Y_{\bbZ_2})$ (which is
the same as F-theory on $Y_{\bbZ_2}$) contains
\begin{align}\label{eq:singlet_multiplicity_pureZ2}
	x_{\bf 1} = 12\,K_B^2 - 8\, K_B \cdot \beta - 4\,\beta^2 
\end{align}
$\bbZ_2$ charged hypermultiplets.  Note that this number agrees with the
number of charge $1$ singlets in the Morrison--Park model \eqref{multMP}.

One can now deform these two geometries in an analogous way to that in section
\ref{sec:resolution_restricted_model}, namely by setting
\begin{equation}\label{eq:tuning_bisection}
  c_1 = 0 \,, \quad c_3 = 0 \,,
\end{equation}
which in section \ref{sec:resolution_restricted_model} engineered a
$\mathbb{Z}_2$ Mordell--Weil group. In this case such a tuning results, as we will momentarily see, 
in $\mathbb{Z}_2$ Mordell--Weil torsion when applied
to the Jacobian fibration, $J(Y_{\mathbb{Z}_2})$, but not in the genus-one
bisection fibration, $Y_{\mathbb{Z}_2}$, as the notion of the Mordell--Weil
group exists only for elliptic fibrations\footnote{However, see
  \cite{Grimm:2015wda} for details of an arithmetic structure on genus-one
fibrations with multi-sections, similar to the Mordell--Weil group.}.
Note that this tuning can not be torically realized in the $\bbP_{112}$ hypersurface \eqref{eq:generic_P112_hypersurface}, as the tuning \eqref{eq:tuning_bisection} does not correspond to removing vertices of the dual polygon (see, e.g., figure 5 in \cite{Morrison:2012ei} for a description in the same notation).

The resulting hypersurface $Y_b$ in the $\bbP_{112}$ fibration,
\begin{align}\label{eq:tuned_bisection_geometry}
	Y_b : \qquad w^2 = c_0\,u^4 + c_2\,u^2\,v^2 + c_4\,v^4 \, ,
\end{align}
remains a smooth genus-one fibration with a bisection, thus we expect 
\begin{align}
	\Sh(J(Y_b)) = \bbZ_2 \, .
\end{align}
For the Jacobian geometry, the tuning $c_1=0$ and $c_3=0$ yields a new elliptic fibration
\begin{equation}\label{eq:weierstrass_jacobian_su2xz2}
  J(Y_b) : \qquad y^2 = x^3 + (- \frac{1}{3}c_2^2 - 4 c_0 c_4)\,x\,z^4 + (- \frac{2}{27}c_2^3 + \frac{8}{3}c_0c_2c_4)\,z^6
  \,, 
\end{equation}
with discriminant
\begin{align}\label{eq:disc_jacobian_su2xz2}
	\Delta_b = c_0\,c_4\,(c_2^2 - 4\,c_0\,c_4)^2 \, .
\end{align}
We can see that this elliptic fibration has, in addition to the zero-section at
\begin{equation}
  [x : y : z] = [1 : 1 : 0] \,,
\end{equation}
also a rational section situated at 
\begin{equation}
  [x : y : z] = \left[\frac{2}{3} c_2 : 0 : 1 \right] \,.
\end{equation}
As discussed in appendix \ref{sec:MW-torsion_Weierstrass}, since such a
rational section has $y = 0$, it necessarily sits at the $\mathbb{Z}_2$ torsion
point of the fibration and thus generates a Mordell--Weil group
\begin{equation}
  {\rm MW}(J(Y_b)) = \mathbb{Z}_2 \,.
\end{equation}
Moreover, it is easily checked that the section passes through the $I_2$ singularity over the discriminant component 
\begin{align}
	c_2^2 - 4\,c_0\,c_4 =0 \, ,
\end{align}
indicating that the corresponding $\mathfrak{su}(2)$ algebra is affected by
the torsional section.  Indeed, a quick glance at the
codimension two loci reveals that the only enhancements along the
$\mathfrak{su}(2)$ divisor are at
\begin{equation}
  \begin{aligned}
    c_0 &= c_2^2 - 4 c_0 c_4 = 0 \, , \\
    c_4 &= c_2^2 - 4 c_0 c_4 = 0 \,,
  \end{aligned}
\end{equation}
both of which support a type $III$ fiber.  The absence of any other
enhancement loci which can support fundamental matter thus further suggests
that the global group structure is actually $SU(2) / \mathbb{Z}_2$.  Away from
the $\mathfrak{su}(2)$ divisor, there is a codimension two $I_2$ enhancement locus at $c_0 =
c_4 = 0$.

These geometric data hints towards an F-theory model with gauge symmetry
\begin{align}
	\frac{SU(2)}{\bbZ_2} \times \bbZ_2 \,.
\end{align}
In the following, we will
provide further evidence that F-theory compactified on $J(Y_b)$ indeed gives rise to
such a field theory.  However, we will also see that the F-theory interpretation, in
particular of the bisection geometry $Y_b$, is much more obscured than in
previously known examples.

\subsection[F-theory on the Jacobian \texorpdfstring{$J(Y_b)$}{J(Yb)}]{F-theory on the Jacobian \boldmath{$J(Y_b)$}}

We now wish to study the physics of the F-theory compactification on this
Jacobian $J(Y_b)$ in more detail.  To this end, we first analyze the $I_2$
singularities above the codimension one locus $c_2^2 - 4\,c_0\,c_4=0$, located
at
\begin{equation}\label{eq:jacobian_I2_sing_codim1}
    c_2^2 - 4 \, c_0\, c_4 = y = 2 c_2\,z^2 - 3x = 0 \,.
\end{equation}
In order to resolve the singularity we shall
first perform a coordinate shift to locate the singularity in the fiber at the
origin
\begin{equation}
  x \rightarrow x + \frac{2}{3}c_2\, z^2 \,,
\end{equation}
yielding the shifted Weierstrass model, which we also refer to as $J(Y_b)$,
\begin{equation}\label{eqn:whatsinalabel}
  J(Y_b) : \qquad y^2 - x \, ((c_2^2 - 4 \, c_0 \, c_4) \, z^4 + 2 \, c_2\, z^2 x + x^2) \, .
\end{equation}
The fibration again has two rational sections
\begin{equation}
  \begin{aligned}[ ]
    [x : y : z ] &= [ 1 : 1 : 0 ] \cr
    [x : y : z ] &= [ 0 : 0 : 1] \,,
  \end{aligned}
\end{equation}
where the first is the zero-section and the latter the $\mathbb{Z}_2$
torsional section.

We can resolve the singularity \eqref{eq:jacobian_I2_sing_codim1} by a blow-up
\begin{equation}
  (x, y, c_2^2 - 4 \,c_0\, c_4 ; \zeta) \,,
\end{equation}
in the notation of \cite{Lawrie:2012gg}. Such a blow-up involves introduction
of a new coordinate $\zeta$ and replacing
\begin{equation}
  x \rightarrow x  \,\zeta \,, \quad y \rightarrow y \, \zeta \,, \quad c_2^2 - 4
  \,c_0\, c_4 \rightarrow A\, \zeta \,,
\end{equation}
where $[x : y : A]$ are now projective coordinates. Performing such a
transformation in the hypersurface (\ref{eqn:whatsinalabel}), followed by the
proper transform, yields the resolved threefold, $\hat{J}(Y_b)$, described by
the complete intersection
\begin{equation}
  \begin{aligned}
    y^2 - x \, (2c_2\,x\,z^2 + x^2\zeta + z^4 \, A) &= 0 \cr
    -c_2^2 + 4\,c_0\,c_4 + \zeta \, A &= 0 \,,
  \end{aligned}
\end{equation}
in a fivefold ambient space with the Stanley--Reisner ideal
\begin{equation}
  \{ x\,y\,z, \zeta \, z , x\,y\,A \} \,.
\end{equation}
The two exceptional divisors associated to the $I_2$ fiber are
\begin{equation}
  \begin{aligned}
    A = 0 \,: &\quad y^2 - x^2 (2\, c_2 \,z^2 + x \, \zeta) = 0 \,, &  &\quad
    c_2^2 - 4\,c_0\,c_4 = 0 \cr
    \zeta = 0 \,: &\quad y^2 - x \, (2 \, c_2 \, x\, z^2 + z^4\, A) = 0 \,, & 
    &\quad c_2^2 - 4\,c_0\,c_4 = 0 \, ,
  \end{aligned}
\end{equation}
where the first corresponds to the affine node, and the second to the Cartan divisor of the $\mathfrak{su}(2)$.
At the codimension two point with $c_2=0$, there is no factorization of the fibral
curves, but the two components intersect in a single point, $x=y=\zeta=0$.
This indeed corresponds to a  type $III$ fiber.  The codimension two
singularities above $c_0 = c_4 = 0$ turn out to be terminal singularities that
cannot be resolved crepantly.  Hence, despite the singularity enhancement from
$I_1$ to $I_2$, locally, the fiber is an $I_1$ with Milnor number $m_P= 1$.
This signals that each point in $c_0 = c_4=0$ supports one hypermultiplet
uncharged under any massless gauge symmetry \cite{Arras:2016evy}.  We have
summarized the singular fibers in table
\ref{tab:singular_fibers_jacobian_su2xz2}.

\begin{table}
	\begin{align*}
		\begin{array}{c|c|c|c}
			\text{Locus} & \text{Fiber Type} & \mathfrak{su}(2)\text{ Rep.} &
      \text{Multiplicity of Hypermultiplets} \\ \hline \rule{0pt}{3ex}
			\{c_0\} & I_1 & - & - \\
			\{c_4\} & I_1 & - & - \\
			\{c_2^2 - 4\,c_0\,c_4\} & I_2 & {\bf 3} & 1 + 6\,K_B^2 \\
			\{c_2\} \cap \{c_0\} & III & - & - \\
			\{c_2\} \cap \{c_4\} & III & - & - \\
			\{c_0\} \cap \{c_4\} & I_1 \, (\text{term.}) & {\bf 1} & -8\,K_B\cdot \beta - 4\,\beta^2
		\end{array}
	\end{align*}
	\caption{Singular fibers and associated matter states of the blown-up Jacobian geometry $\hat{J}(Y_b)$. The singularities over $\{c_0\} \cap \{c_4\}$ are terminal and thus not resolved.}\label{tab:singular_fibers_jacobian_su2xz2}
\end{table}

Thus, we conclude that F-theory on the partially resolved Jacobian geometry
$\hat{J}(Y_b)$ has an $SU(2)/\bbZ_2$ gauge symmetry without any
localized matter.  The gauge anomalies are straightforwardly checked to be
canceled.  To verify the gravitational anomaly cancellation, we can compute
the Euler characteristic of $\hat{J}(Y_b)$ with the procedure laid out in
section \ref{sec:grav_anomaly_specialized_model}.  As explained in
\cite{Arras:2016evy}, the contribution of the terminal singularities are
accounted for correctly if we treat the fibers as $I_1$ curves with $\chi_{\rm
top}=1$.  Thus we obtain the following contributions:
\begin{align}\label{eq:chi_su2xz2}
	\begin{split}
		\text{Codimension Two}: \qquad & 3\,[c_2] \cdot ([c_0] + [c_4]) + [c_0] \cdot [c_4] \\
		\text{Codimension One}: \qquad & - ([c_0] + K_B)\cdot [c_0] - [c_0] \cdot ([c_2] + [c_4]) \\
		& - ([c_4] + K_B) \cdot [c_4] - [c_4] \cdot ([c_0] + [c_2]) \\
		& + 2 \times \left( - ([c_2^2] + K_B) \cdot [c_2^2] - [c_2] \cdot ([c_0] + [c_4]) \right)\\
		\Longrightarrow \qquad & \chi_{\rm top}(\hat{J}(Y_b)) = -36\,K_B^2 -
    8\,K_B\cdot \beta - 4\, \beta^2 \, .
	\end{split}
\end{align}
In the presence of terminal singularities, the Euler characteristic satisfies \cite{Arras:2016evy}
\begin{align}
	\chi_{\rm top} = 2 + 2\,b_2 - b_3 \, ,
\end{align}
where the second Betti number $b_2 = 1 + h^{1,1}(B) + {\rm rk}(\fkg) = 11 -
K_B^2 + {\rm rk}(\fkg)$ still satisfies the Shioda--Tate--Wazir formula.
As spelled out in \cite{Arras:2016evy}, the non-localized uncharged hypermultiplets are counted by
\begin{align}
	n_{H, {\rm n.l.}}^0 = \frac{1}{2} \left( b_3 - \sum_P m_P \right) \, ,
\end{align}
whereas the localized uncharged hypermultiplets are counted by the points $P \in B$
with terminal singularities, weighted by the associated Milnor number $m_P$,
\begin{equation}
  n_{H, {\rm l.}}^0 = \sum_P m_P \,.
\end{equation}
In our case, there are $[c_0] \cdot [c_4]$ points with terminal singularities
having $m_P=1$.
We thus have in total
\begin{align}\label{eq:neutral_hypers_Jacobian_su2xz2}
	n_H^0 = \frac{1}{2} \left( b_3 + \sum_P m_P \right) = \frac{1}{2} \left( 2 +
  2\,b_2 - \chi_{\rm top} + [c_0]\cdot [c_4] \right) = 13 + 17\,K_B^2 \, .
\end{align}
The only charged hypermultiplets come from the (charged) states of the
$\mathfrak{su}(2)$ adjoint representation, giving $n_H^c = 2 + 12K_B^2$ (see table \ref{tab:singular_fibers_jacobian_su2xz2}).
With three vector multiplets from the $\mathfrak{su}(2)$ gauge fields, the gravitational anomaly,
\begin{align}
	n_H^0 + n_H^c - n_V + 29\,n_T = 15 + 29\,K_B^2 - 3 + 29\,(9 - K_B^2) = 273 \, ,
\end{align}
cancels perfectly.

In the Jacobian description, we can only see the massless $\mathfrak{su}(2)$
gauge symmetry at the level of divisors.  The presence of terminal
singularities however signals some broken gauge symmetry under which the
localized matter are charged.  Since the Jacobian arises as the Jacobian of a geometry with
a bisection, a natural proposal is that there is a $U(1)$ broken to a
$\bbZ_2$.  Such a discrete remnant would manifest itself as a non-trivial
torsion subgroup of $H^3(J(Y_b), \bbZ)$ \cite{Braun:2014oya,
Mayrhofer:2014laa}, which, however, is notoriously difficult to determine
explicitly.  We will refrain ourselves from attempting the necessary
computation, and instead, in the next subsection, give evidence for the
presence of the additional $\bbZ_2$ discrete symmetry based on the consistency
of Higgsing chains.

\subsection{Matching the Spectrum via Higgsing}

Any complex structure deformation of the geometry that modifies the gauge
algebra and the multiplicities of the matter hypermultiplets  corresponds to a
field theoretic Higgsing that modifies the algebra and fields in the same way.
Concretely, in our case, we have in mind a sequence of complex structure
deformations with corresponding 6D field theory Higgsings:
\begin{equation}\label{diagram}
	\begin{tikzcd}[row sep = normal, column sep= normal
    ]
		\text{Geometry:} & J(\hat{Y})  &\arrow[l, "c_4 \rightarrow \frac{b^2}{4}"'] J(Y_b) & \arrow[l, "c_{3} \rightarrow 0", "c_1\rightarrow 0"'] J(Y_{\bbZ_2}) \\
		\text{Field theory:} & \frac{SU(2)_A}{\bbZ_2} \times SU(2)_B \arrow[r, "\text{(a)}"] & \frac{SU(2)}{\bbZ_2} \times H \arrow[r, "\text{(b)}"] & \bbZ_2
	\end{tikzcd}
\end{equation}
The geometry $J(\hat{Y})$ giving rise to an $SU(2)_A/\bbZ_2 \times SU(2)_B$
theory is the Weierstrass model of the specialized $\bbZ_2$-tuned
Morrison--Park model discussed in section
\ref{sec:resolution_restricted_model}.  Specifically, its spectrum is
summarized in \eqref{tab:restricted_example_localized_fiber_curves}.  On the
other side, the geometry $J(Y_{\bbZ_2})$ is defined in
\eqref{eq:standard_bisection_model_jacobian} and is well-known to give rise to
a theory with only $\bbZ_2$ symmetry.  The possible discrete symmetry $H$ in
the middle must fit into the chain of Higgsing, where the individual steps (a)
and (b) must be compatible with the massless $SU(2)/\bbZ_2$ in the middle as
well as the charged and uncharged spectra of the theories on the end of that
chain.

\subsubsection*{Higgsing step (a)}

It turns out that the only compatible Higgsing (a) is a two-step breaking
process, by first breaking $SU(2)_B \rightarrow U(1)$ with an adjoint
hypermultiplet $(\bf 1, 3)$, and then break the $U(1)$ with the ${\bf 1}_{2}$
states arising as remnants of the remaining $(\bf 1,3)$ hypermultiplets.  This
breaks the $U(1)$ to a $\bbZ_2 = H$.  Note that for breaking the $U(1)$ in a
D-flat manner, one needs two hypermultiplets of ${\bf 1}_2$ singlets.  Since
this breaks $SU(2)_B$ to a discrete group, all three gauge bosons acquire a
mass, and hence `eat up' three hypermultiplets of the Higgs field according to
Goldstone's theorem.\footnote{In this counting, the adjoint ${\bf 3}$ of
$\mathfrak{su}(2)$ contains three hypermultiplets.} Thus, $2 \times \#{(\bf
1,3)} - 3$ additional uncharged hypermultiplets arise from $(\bf 1,3)$ states
in the Higgsing process (a).  Note that the prefactor is $2$, because one
hypermultiplet of each $(\bf 1,3)$ is uncharged under the Cartan of $SU(2)_B$
and hence already accounted for in the number \eqref{eq:uncharged_hypers_su2xsu2} of uncharged hypermultiplets
of the $SU(2)/\bbZ_2 \times SU(2)$ theory.  Denoting the $\bbZ_2$ even/odd charges
by subscripts with $0$/$1$, the other representations
in table \ref{tab:spectrum_restricted_model} decompose as
\begin{align}\label{eq:decomposition_su2xsu2-su2xz2}
	\begin{array}{c|c}
		\mathfrak{su}(2)^{\oplus 2} \rightarrow \mathfrak{su}(2) \oplus \bbZ_2 & \text{multiplicities}  \\ \hline \rule{0pt}{3ex}
		({\bf 3,1}) \rightarrow {\bf 3}_0 & 1 + 2K_B^2 - 2K_B\cdot \beta \\
		({\bf 1,2}) \rightarrow 2 \times {\bf 1}_1 & -4K_B\cdot \beta- 2\beta^2 \\
		({\bf 3 \oplus 1,2}) \rightarrow 2 \times {\bf 3}_1 \oplus 2 \times {\bf
    1}_1 & 2 K_B^2 + K_B \cdot \beta
	\end{array}
\end{align}

To match this Higgsed spectrum with that of F-theory on $J(Y_b)$, we note that
on $J(Y_b)$, the geometric counting in the previous section does not
distinguish between states of different  $\bbZ_2$ charge, e.g.,
$\mathfrak{su}(2)$ singlets are all counted as uncharged hypermultiplets.
In that case, the matching of the charged spectrum is straightforward, as it becomes
just counting the number of $\mathfrak{su}(2)$ adjoints after the Higgsing.
From \eqref{eq:decomposition_su2xsu2-su2xz2}, we easily spot the $1 + 6K_B^2$
adjoint representations needed to match the geometric counting in table
\ref{tab:singular_fibers_jacobian_su2xz2}.  Furthermore, we can also match the
uncharged spectrum.  Explicitly, we obtain additional uncharged hypermultiplets from
the ${\bf 1}_1$ and, importantly, also from the $\bbZ_2$ charged adjoints
${\bf 3}_1$, where the state without Cartan charge is also an uncharged
hypermultiplet in the geometric counting
\eqref{eq:neutral_hypers_Jacobian_su2xz2}.  Note that even though the ${\bf
3}_0$ also contains a uncharged hypermultiplet, we do not have to include them
in the counting of additional uncharged hypermultiplets, because they were
already accounted for in the $SU(2)/\bbZ_2 \times SU(2)$ model
\eqref{eq:uncharged_hypers_su2xsu2}.  In total, we then have
\begin{align}
	2 \times \#({\bf 1,3}) -3 + 2 \times \#({\bf 1,2}) + 4 \times \# ({\bf 3
  \oplus 1, 2}) = 10 K_B^2 - K_B\cdot \beta - 3\beta^2 - 1
\end{align}
additional uncharged hypermultiplets arising in the Higgsing (a), which
together with the already present uncharged hypermultiplets
\eqref{eq:uncharged_hypers_su2xsu2} in the $SU(2)/\bbZ_2 \times SU(2)$ phase precisely matches the number
\eqref{eq:neutral_hypers_Jacobian_su2xz2} computed geometrically for
$\hat{J}(Y_b)$.

\subsubsection*{Higgsing step (b)}

The above Higgsing leads to an $SU(2)/\bbZ_2 \times \bbZ_2$ theory with the following charged spectrum
\begin{align}\label{eq:charged_matter_multiplicities_su2xz2}
	\begin{array}{c|c}
		\mathfrak{su}(2) \oplus \bbZ_2 \text{ Rep} & \text{Multiplicity of
    hypermultiplets} \\ \hline \rule{0pt}{3ex}
		{\bf 3}_0 & 1 + 2K_B^2 - 2K_B\cdot \beta \\
		{\bf 3}_1 & 4K_B^2 + 2K_B \cdot \beta \\
		{\bf 1}_1 & 4K_B^2 - 6K_B\cdot \beta - 4\beta^2
	\end{array}
\end{align}
where the subscript denotes $\bbZ_2$ charge.  In order to higgs this to a
theory with just a $\bbZ_2$ discrete symmetry, corresponding to F-theory on
$J(Y_{\bbZ_2})$, we again need a two-step Higgsing process.  First we give a
vacuum expectation value to a hypermultiplet in the ${\bf 3}_0$ representation, breaking
$SU(2)/\bbZ_2 \times \bbZ_2$ to $U(1) \times \bbZ_2$.  Under this breaking,
the $\bbZ_2$ charged adjoints ${\bf 3}_1$ decompose into singlets which are
charged under both $U(1)$ and $\bbZ_2$.  Higgsing such a singlet then further
breaks the gauge symmetry to a diagonal $\bbZ_2$.  Explicitly, we obtain the
following decomposition:
\begin{equation}\label{eq:field_theory_breaking_(b)}
	\begin{array}{ccccc}
		SU(2) /\bbZ_2 \times \bbZ_2 & \stackrel{\langle {\bf 3}_0 \rangle}{\longrightarrow} & U(1) \times \bbZ_2 & \stackrel{\langle {\bf 1}_{(1,1)} \rangle}{\longrightarrow} & \bbZ_2 \\ \hline \rule{0pt}{3ex}
		{\bf 3}_0 & \longrightarrow & 2 \times {\bf 1}_{(1,0)} \oplus {\bf 1}_{(0,0)} & \longrightarrow & 2 \times {\bf 1}_1 \oplus {\bf 1}_0 \\
		{\bf 3}_1 & \longrightarrow & 2 \times {\bf 1}_{(1,1)} \oplus {\bf 1}_{(0,1)} & \longrightarrow & 2 \times {\bf 1}_0 \oplus {\bf 1}_1 \\
		{\bf 1}_1 & \longrightarrow & {\bf 1}_{(0,1)} & \longrightarrow & {\bf 1}_{1}
	\end{array}
\end{equation}
It is straightforward to sum up the contributions to the singlets with
$\bbZ_2$ charge $1$, which, with the multiplicities in
\eqref{eq:charged_matter_multiplicities_su2xz2}, yields
\begin{align}
	2 \times (\# ({\bf 3}_0) - 1) + \# ({\bf 3}_1) + \# ({\bf 1}_1) = 12K_B^2 -
  8K_B\cdot \beta - 4\beta^2 \, ,
\end{align}
which is the number (\ref{eq:singlet_multiplicity_pureZ2}) of $\bbZ_2$ charged singlets in the F-theory
compactification on $J(Y_{\bbZ_2})$.

\subsection[F-theory on the Bisection Model \texorpdfstring{$Y_b$}{Yb}]{F-theory on the Bisection Model \boldmath{$Y_b$}}\label{sec:non_F-theory}

Thus far, we have analyzed the 6D F-theory compactification on the Jacobian
fibration $J(Y_b)$.  Requiring the two complex structure deformations \eqref{diagram}, that
connect $J(Y_b)$ with the $\bbZ_2$ torsion-enhanced Morrison--Park model
$J(\hat{Y})$ and with the standard $\bbZ_2$ model $J(Y_{\bbZ_2})$, to be
consistent with a field theoretic Higgsing process constrains the 6D theory to
be an $SU(2)/ \bbZ_2 \times \bbZ_2$ gauge theory.  This conclusion is
further supported by the fibration structure of $J(Y_b)$, which has a
codimension one locus of $I_2$ fibers, a $\bbZ_2$ torsional section, and
terminal singularities in codimension two.

Based on observations made throughout the literature, one expects that an $n$-section geometry $Y$ should give rise to the same F-theory compactification
as its Jacobian $J(Y)$.  Specifically, we know \cite{Braun:2014oya,
Mayrhofer:2014laa, Arras:2016evy} that we can uplift the 5D M-theory
compactification on $J(Y)$ through the standard M-/F-duality to a 6D theory --
\textit{the} F-theory compactification on $J(Y)$ -- despite the presence of
codimension two terminal singularities.  In all known examples, the 5D
M-theory compactification on the $n$-section geometry $Y$ could be identified as
a circle reduction of the 6D F-theory with a non-zero flux of 
a massive $U(1)$, which gauges the $\mathbb{Z}_n$, along the circle \cite{Anderson:2014yva,
Garcia-Etxebarria:2014qua, Mayrhofer:2014haa, Cvetic:2015moa}.  The field
theoretic consequence of this flux is that in 5D, the $\bbZ_n$ symmetry
is identified as a subgroup of the Kaluza--Klein $U(1)$, which now
manifests itself as the $U(1)$ in M-theory dual to the $n$-section divisor.
Crucially, the rank of the massless gauge symmetry in the 5D M-theory
compactification is the same for both the Jacobian and the bisection geometry.

This situation persists in models where the 6D theory includes non-abelian
gauge algebras $\fkg$ \cite{Braun:2014oya, Anderson:2014yva,
Garcia-Etxebarria:2014qua, Mayrhofer:2014haa, Braun:2014qka, Lin:2015qsa,
Oehlmann:2016wsb}.  In this setup, on the 5D Coulomb branch of either the Jacobian or
the $n$-section geometry $Y_b$, one must have $\text{rk}(\fkg) + \#(\text{indept.
$n$-sections})$ independent $U(1)$s.  Geometrically, this means that we need
$h^{1,1}(Y) = h^{1,1}(B) + \text{rk}(\fkg) + \#(\text{indept. $n$-sections})$.  This
however is not true for the case at hand!  
The deformation from $Y_{\bbZ_2}$ to $Y_b$ is a smooth deformation, and thus
one has the same number, $h^{1,1}(Y_{\bbZ_2})$, of divisors
\begin{equation}
  h^{1,1}(Y_b) = h^{1,1}(Y_{\bbZ_2})= h^{1,1}(B) + 1 \,.
\end{equation}
  This does not match the number of divisors in the Jacobian model $J(Y_b)$,
  where the tuning is not a smooth deformation of $J(Y_{\mathbb{Z}_2})$, but
  introduces new singularities, and thus one finds that 
\begin{equation}
  h^{1,1}(J(Y_b)) = h^{1,1}(B) + 2 \,,
\end{equation}
where the additional divisor corresponds to the Cartan divisor of the
$\mathfrak{su}(2)$.

This mismatch obscures a direct interpretation of F-theory on $Y_b$ through the
standard M-/F-duality.  The geometry seems to preclude
both the Cartan $U(1)$ of the $SU(2)/\bbZ_2$ and the
Kaluza--Klein $U(1)$ from being realized independently on the 5D Coulomb branch.  It would be interesting
to understand, if one can make sense of the geometry in such a way that we can
still interpret M-theory compactified on $Y_b$ and $J(Y_b)$ as different
circle reductions of the same 6D theory, or if there are inherent differences
to previously studied models, such that F-theory on the genus-one fibration
$Y_b$ involves some particular subtleties.

Furthermore, the consistency of the field theory Higgsing required the
existence of both $\bbZ_2$ charged and uncharged adjoints of the
$\mathfrak{su}(2)$.  Surprisingly, both charged and uncharged adjoints
\eqref{eq:charged_matter_multiplicities_su2xz2} account together for the
multiplicity of deformation adjoints (see table
\ref{tab:singular_fibers_jacobian_su2xz2}) of the $\mathfrak{su}(2)$ divisor
in the Jacobian geometry $J(Y_b)$. To our knowledge, this is the first example
where not all deformation modes of the same non-abelian gauge divisor carry
the same charges.  This conclusion is based on field theory arguments, which
we believe should have a counterpart in geometry. In the Jacobian geometry
$J(Y_b)$, the discrete symmetry is encoded in torsional three-cycles
\cite{Mayrhofer:2014laa} and is difficult to study directly. A better
understanding of the bisection geometry may allow one to read off the discrete
charges, including of non-localized adjoints, more directly.

\section{Conclusions and Future Directions}

In this paper, we have put forth a procedure to construct Weierstrass models of elliptic fibrations with a torsional Mordell--Weil generator that can be deformed to a free rational section.
This procedure is exemplified by determining the Weierstrass model \eqref{eq:Weierstrass_functions_generic_solution}, which is birationally equivalent to any such elliptic fibration with a $\bbZ_2$ torsional section.
When this Weierstrass model
is a Calabi--Yau threefold, the F-theory
compactification to 6D yields a field theory with gauge group
\begin{equation}
  \frac{SU(2)_A \times SU(4)}{\mathbb{Z}_2} \times SU(2)_B \, ,
\end{equation}
where one of the notable features is that the quotient does not act on
every non-abelian factor of the gauge algebra.

Furthermore, we have found that the solution exhibits a singular discriminant component hosting the $\mathfrak{su}(2)_A$ gauge algebra.
At the self-intersection locus, where also the $\mathfrak{su}(2)_B$ divisor passes through, there is matter in the $({\bf 2 \otimes 2, 1, 2}) = ({\bf 3, 1, 2}) \oplus ({\bf 1, 1,2})$ representation.
While we have not attempted a full resolution of the generic model, we have identified a subsolution with gauge group $SU(2)_A/\bbZ_2 \times SU(2)_B$ exhibiting the same feature, and which also allows for a toric resolution as described in section \ref{sec:resolution_restricted_model}.
In this global resolution one observes that this matter is realized by an reduced $I_0^*$ fiber over the singular point of the discriminant component.

While we in principle have solved the $\bbZ_2$ torsional condition for any elliptic fibration with non-trivial Mordell--Weil rank, it remains a difficult problem to study the associated F-theory compactification if the Weierstrass fibration of the form \eqref{eq:Weierstrass_functions_generic_solution} is not Calabi--Yau.
In particular, this complication may arise if we want to study analogous gauge enhancements of F-theory models with higher rank Mordell--Weil group or higher charge singlets.
With the plethora of explicit F-theory models with rank $\geq 2$ Mordell--Weil group \cite{Borchmann:2013jwa, Borchmann:2013hta, Cvetic:2013nia, Cvetic:2013jta, Cvetic:2013qsa, Cvetic:2015ioa, Krippendorf:2015kta}, an obvious extension would therefore be to solve the torsion condition directly in these constructions.
For example, in models with multiple $U(1)$s, tuning one or more rational sections to be torsional might produce non-abelian gauge algebras with more intricate global structures.
Furthermore, finding generic solutions for Mordell--Weil torsion $\Gamma \neq \mathbb{Z}_2$ in general
may be more than just an exercise in commutative algebra, and may give
insights into new F-theory physics.

One may have hoped that by tuning the section to be torsional, the resulting gauge enhancement could also have produced higher dimensional representations, similar to constructions where one collides two or more sections \cite{Morrison:2012ei, Morrison:2014era, Cvetic:2015ioa, Klevers:2016jsz}.
There, the higher $U(1)$ charges become the Cartan charges of higher dimensional non-abelian representations after the enhancement.
On the other hand, when one tunes the section of the so-called
$U(1)$-restricted Tate model \cite{Grimm:2010ez} to be $\bbZ_2$ torsional, one finds \cite{Mayrhofer:2014opa} that the charge 1 singlets (which are the only charged singlets of the restricted Tate model) enhances to $\mathfrak{su}(2)$ adjoints with (highest) Cartan charge 2.
Naively, one might have expected that the analogous tuning of a $U(1)$ model with charge 2 singlets would result in an $\mathfrak{su}(2)$ model with a ${\bf 5}$ representation that has highest Cartan charge 4.
However, our generic solution does not exhibit such higher dimensional representations, but instead a higher rank gauge group, whose breaking then yields the higher charged singlets.
If we seek to restrict the generic solution so that the enhancement is rank
preserving, or $U(1) \rightarrow SU(2)/\mathbb{Z}_2$, then one finds that one
must turn off the charge 2 locus in the $U(1)$ model.
This finding is consistent with the recent ``swampland conjecture'' \cite{Klevers:2017aku}, which forbids higher dimensional representations such as the ${\bf 5}$ of $\mathfrak{su}(2)$ in F-theory compactifications.
It would be interesting to analyze if in models with $U(1)$ charge $> 2$ \cite{Klevers:2014bqa, Cvetic:2015moa, Morrison:2016xkb}, whose Morrison--Park model has ``tall'' sections and thus are non-Calabi--Yau \cite{Morrison:2016xkb}, a torsional enhancement would generate novel higher dimensional representations.

Finally, we have studied a related deformation process of a Weierstrass model that arises as the Jacobian $J(Y_{\bbZ_2})$ of the bisection geometry $Y_{\bbZ_2}$, whose F-theory compactification gives rise to a $\bbZ_2$ symmetry \cite{Braun:2014oya, Anderson:2014yva, Garcia-Etxebarria:2014qua, Mayrhofer:2014haa, Mayrhofer:2014laa, Morrison:2014era}.
In section \ref{sec:rosie}, we have seen that the deformed Jacobian $J(Y_b)$ exhibits a $\bbZ_2$ torsional section that comes with the massless $SU(2)/\bbZ_2$ gauge symmetry.
At the same time, this Jacobian also has terminal singularities, which we argued field theoretically corresponds to localized singlets charged under a discrete $\bbZ_2$ gauge symmetry.
However, the field theory is rather peculiar, e.g., it contains $\mathfrak{su}(2)$ adjoints with different $\bbZ_2$ charges, which appear nevertheless not to be localized in geometry.
From previous examples with discrete symmetries in the literature, one might have hoped to understand $\bbZ_2$ charged matter better in the corresponding deformed bisection geometry $Y_b$.
However, the interpretation of the bisection geometry within the F-theory context only raises more questions.
In particular, the genus-one fibration $Y_b$ lacks independent divisors giving rise to the Cartan of $\mathfrak{su}(2)$ and the Kaluza--Klein $U(1)$ that would be necessary to straightforwardly uplift M-theory on $Y_b$ to \textit{the} F-theory defined on $J(Y_b)$.
Since there are by now a multitude of multi-section models in the literature that can consistently incorporate non-abelian gauge symmetries, it might possible, through a refined definition of F-theory on multi-section geometries such as $Y_b$, to resolve some of these puzzles.
We hope to shed some light on this issue in the future \cite{bisection_to_appear}.

\subsection*{Acknowledgements}

\noindent We thank Thomas Grimm, Christoph Mayrhofer, Paul Oehlmann, Sakura
Sch\"afer-Nameki, Wati Taylor, and Yinan Wang for discussions. We especially thank
Timo Weigand for discussion and collaboration on a related topic.
M.C., C.L., and L.L.~thank CERN for hospitality during the final stages of this
work. F.B.~is supported by the Heidelberg Graduate School for Fundamental Physics. 
M.C.~and L.L.~are supported by DOE Award DE-SC0013528. M.C.~further
acknowledges the support by the Fay R.~and Eugene L.~Langberg Endowed Chair
and the Slovenian Research Agency.
C.L.~is supported by the DFG under Grant TR33 ``The Dark Universe'' and
under GK 1940 ``Particle Physics Beyond the Standard Model''.

\appendix

\section{Mordell--Weil Torsion in Weierstrass Models}\label{sec:MW-torsion_Weierstrass}

Rational points on a smooth elliptic curve $\CE$ form an abelian group, the so-called Mordell--Weil group, see e.g.~\cite{MR2514094}. After briefly reviewing the group law, we show how to find points with $\mathbb{Z}_2$ and $\mathbb{Z}_3$ torsion.

\begin{figure}[p]
\centering
	\def\svgwidth{.4\vsize}
	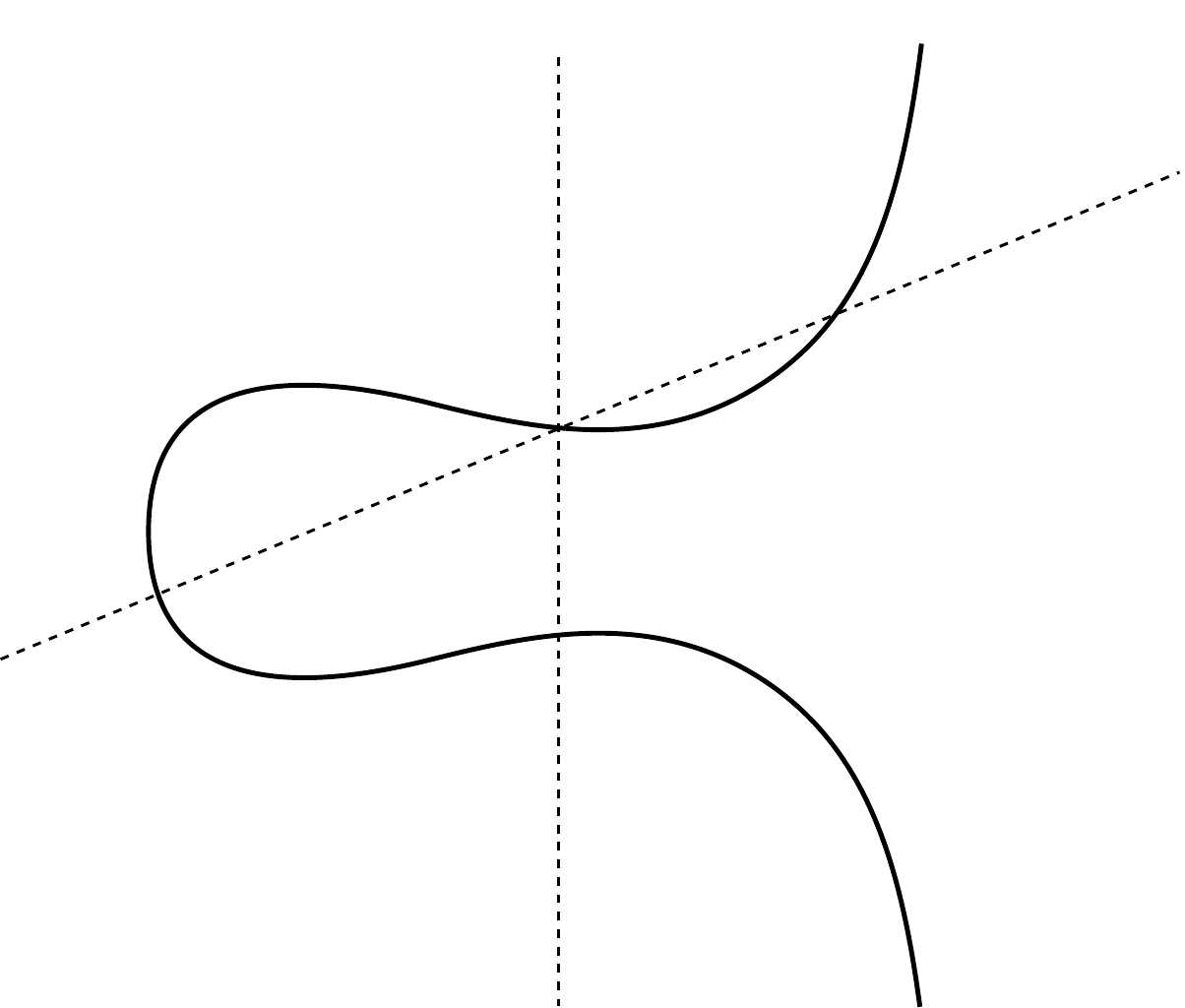
	\caption{Geometric construction of the Mordell--Weil group law. Each dashed line marks three points on the elliptic curve (solid curve) that add up to zero under the group law. The rational points $A,B,C$ satisfy $A \boxplus B = C$.}
	\label{fig:elliptic_group_law}
	\vspace{2cm}
	\def\svgwidth{.4\hsize}
	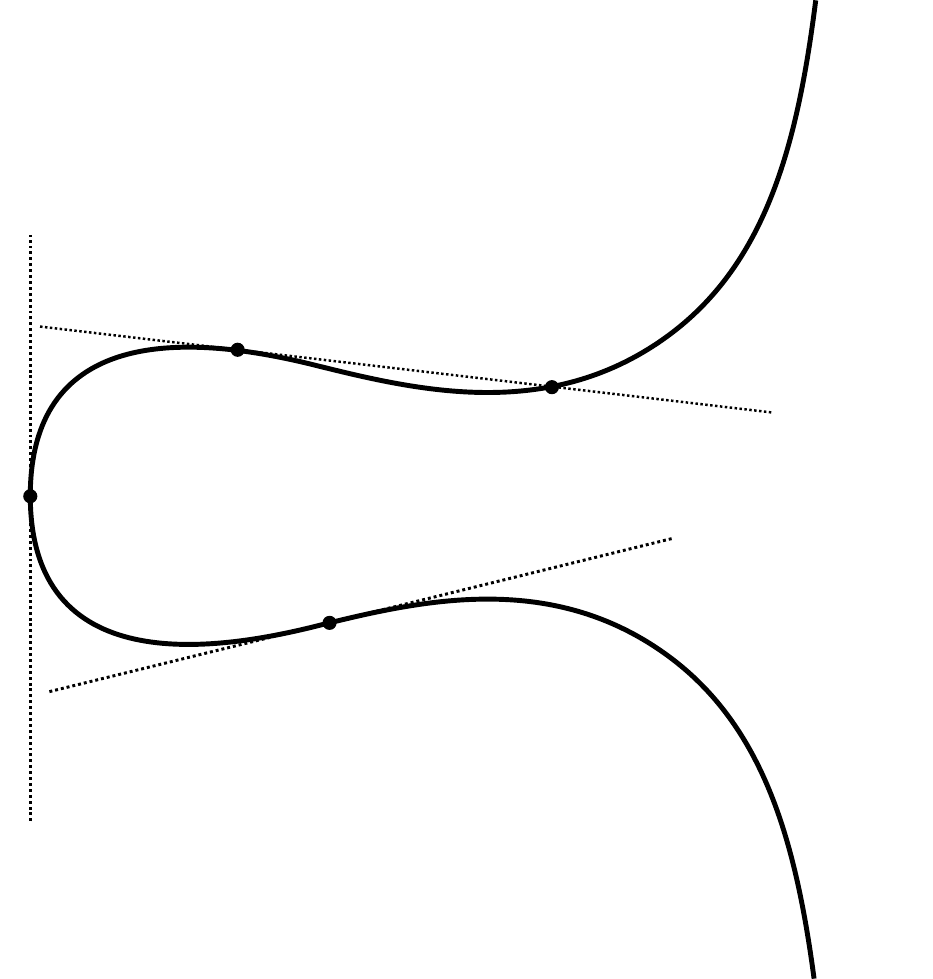
	\caption{A tangent line (dotted) through a point $Q$ intersecting $\CE$ at
  $-R$ corresponds to the Mordell--Weil relation $Q \boxplus Q \boxplus (-R) =
O \Leftrightarrow 2Q = R$. If the tangent line at $Q_2$ is vertical, it
intersects $\CE$ only at infinity, so $2Q_2 \boxplus O= O$. A tangent line at
an inflection point $Q_3$ can be viewed as the limit of taking $Q \rightarrow
Q_3$, which also sends $R$ to $Q_3$. So $2 Q_3 = -Q_3 \Leftrightarrow 3 Q_3 =
O$.}
	\label{fig:MW_torsion}
\end{figure}

In the inhomogeneous Weierstrass form,
\begin{align}\label{eq:weierstrass_inhomogenous}
	P^{\rm inh}_W := -y^2 + x^3 + f\,x + g = 0\, ,
\end{align}
the zero element $O$ of the group is the point at infinity.
Given two rational points $A,B \in \CE$, the straight line in the $(x,y)$-plane through them intersects $\CE$ in a third rational point $C$ (possibly equal to $O$).
Denoting the group action by $\boxplus$, the points satisfy $A \boxplus B \boxplus C = O$.
This geometric construction of the group law is depicted in figure \ref{fig:elliptic_group_law}.
Adding a point $A$ to itself can be seen as the limit of sending the point $B$ to $A$, in which case the line through $A$ and $B$ becomes the tangent at $A$.

Torsional points of order $n$ under the Mordell--Weil group law are points $Q_n$ such that 
\begin{equation}
	\underbrace{Q_n \boxplus Q_n \boxplus ... \boxplus Q_n}_{n\text{ times}} = n\,Q_n = O.
\end{equation}
For $n=2,3$, this relations can be visualized fairly easily, as depicted in figure \ref{fig:MW_torsion}.
In particular, we see that a rational point with a vertical tangent is a $\bbZ_2$ torsional point.
Likewise, a rational inflection point is a $\bbZ_3$ torsional point.

\subsection[\texorpdfstring{$\bbZ_2$}{Z2} Torsion]{\boldmath{$\bbZ_2$} Torsion}\label{app:MW_torsion_2}

In the following, we would like to argue that on a smooth elliptic curve in the inhomogeneous Weierstrass form \eqref{eq:weierstrass_inhomogenous}, a rational point on the curve with $y = 0$ constitutes a $\bbZ_2$ element under the group law.
First note that clearly, any vertical line (i.e., with $x = \text{const.}$) can intersect the curve at most twice at finite values of $x,y$.
The two points are inverse to each other under the Mordell--Weil group law (see figures \ref{fig:elliptic_group_law}) and \ref{fig:MW_torsion}.
Let us now parametrize the curve \eqref{eq:weierstrass_inhomogenous} as two branches with $(x(t), y(t) ) = (t, \pm \sqrt{t^3 + f\,t +g})$, where the sign depends on the branch.
In that parametrization, it is also easy to compute the slope of the curve:
\begin{align}\label{eq:slope_weierstrass_curve}
	\frac{{\rm d} y}{{\rm d} x} = \frac{{\rm d} y(t)}{{\rm d} t} = \pm \frac{3\,t^2 + f}{2 y(t)} = \pm \frac{3\,x^2 + f}{2 y} \, .
\end{align}
Because the curve is smooth by assumption, we know that $P_W$ and its derivative
\begin{align}
	{\rm d} P^{\rm inh}_W = -2\,y \,{\rm d}y + (3 x^2 + f) \, {\rm d}x
\end{align}
cannot vanish simultaneously. In particular, this means that at a point with $y = 0$, the numerator $3x^2 + f$ in \eqref{eq:slope_weierstrass_curve} cannot be zero.
In turn, this means that the slope of elliptic curve \eqref{eq:weierstrass_inhomogenous} must be infinite at a point $Q$ with $y=0$.
Thus, a tangent line at such a point is vertical and intersects the elliptic curve again only at infinity.
From the above discussion (see also figure \ref{fig:MW_torsion}), $Q$ is $\bbZ_2$ torsional.

\subsection[\texorpdfstring{$\bbZ_3$}{Z3} Torsion]{\boldmath{$\bbZ_3$} Torsion}\label{app:MW_torsion_3}

A $\bbZ_3$ torsional point $Q_3$, i.e., $3 Q_3 = 2Q_3 \boxplus Q_3 = 0$, implies that the tangent at $Q_3$ can intersect the elliptic curve only in $Q_3$ again. This is only possible if $Q_3$ is an inflection point, in which case the intersection multiplicity (in the sense of B\'{e}zout's theorem) of the tangent with the curve is 3.
Alternatively, one can also view the inflection point $Q_3$ as the limit of approaching two points $Q$ and $R$ that satisfy $2Q = R$, see figure \ref{fig:MW_torsion}.

From the expression for the slope \eqref{eq:slope_weierstrass_curve}, we can easily determine its second derivative:
\begin{equation}
	\frac{{\rm d}^2 y}{{\rm d}x^2} = \frac{\rm d}{{\rm d} x} \underbrace{\left( \pm \frac{3\,x^2 +f}{2 y(x)}\right)}_{y'(x)} = \pm \frac{6\,x\,y(x) - (3\,x^2 + f)\,y'(x)}{2\,y^2(x)}
\end{equation}
Then, $Q$ is an inflection point if it satisfies the condition
\begin{equation}
	0 \stackrel{!}{=} 6\,x_Q\,y_Q - (3\,x_Q^2 + f)\,y'(x)|_Q \quad \Longleftrightarrow \quad 12 \, x_Q \, y_Q^2 \stackrel{!}{=} \pm (3\,x_Q^2 + f)^2.
\end{equation}
The sign ambiguity simply reflects the fact that a smooth elliptic curve has two inflection points, or equivalently, the Weierstrass equation is symmetric under $y \leftrightarrow -y$; we will stick with $+$ for definiteness.
Furthermore, note that this relationship is derived from the inhomogeneous form \eqref{eq:weierstrass_inhomogenous} of the Weierstrass equation.
To obtain an expression that is valid for an elliptic fibration, one needs to projectivize the $(x,y)$-plane to $\bbP_{231}$, i.e., including the appropriate factors of $z$.\footnote{This was not necessary for the discussion of $\bbZ_2$ torsion, because there, we were only interested in the denominator of \eqref{eq:slope_weierstrass_curve}, which is just $y$ and does not receive any factors of $z$ through projectivization.}
Thus, the condition for a rational point $Q$ in an elliptic fibration to be $\bbZ_3$ torsional is
\begin{align}\label{eq:condition_for_Z3-torsion}
	(3\,x_Q^2 + f\,z_Q^4)^2 - 12 \, x_Q \, y_Q^2 \stackrel{!}{=} 0 \, .
\end{align}

\section[Gauge Enhancement via \texorpdfstring{\boldmath{$\mathbb{Z}_3$}}{Z3} Torsion]{Gauge Enhancement via \boldmath{$\mathbb{Z}_3$} Torsion}\label{sec:Z3_unhiggsing}

In this appendix we explore the geometry where the section of the elliptic
fibration is located a point of $\mathbb{Z}_3$ torsion. After finding a
simplified solution to the tuning condition and the associated F-theory spectrum, we study possible Higgsings back to
a $U(1)$ model and match the multiplicities.

\subsection[Deforming to \texorpdfstring{$\mathbb{Z}_3$}{Z3} Torsion]{Deforming to \boldmath{$\mathbb{Z}_3$} Torsion}

Starting with the Weierstrass coordinates of the section in the Morrison--Park
model \eqref{eqn:U1WS}, we can apply the same procedure as in section
\ref{sec:Z2enh} to the point of $\mathbb{Z}_3$ torsion. In that case the
tuning condition \eqref{eq:condition_for_Z3-torsion} becomes
\begin{align}\label{eq:Z3-tuning_for_MP}
\begin{split}
	3\,c_3^8 = \, & b^2 \left[ b^{10}\,c_0^2 + b^{8}\,(-2\,c_ 0\,c_ 1\,c_ 3-2\,c_ 0\,c_ 2^2+2\,c_1^2\,c_ 2) + b^6\,(8\,c_0\,c_2\,c_3^2-2\,c_1^2\,c_3^2-6\,c_1\,c_2^2\,c_ 3+c_ 2^4) \right. \\
	& \left. + \, b^4\,(12\,c_ 1\,c_ 2\,c_ 3^3-6\,c_ 0\,c_3^4) + b^2\,(-6\,c_ 1\,c_ 3^5-6\,c_2^2\,c_ 3^4) + 8\,c_ 2\,c_3^6 \right] \, .
\end{split}
\end{align}
To find solutions to this equation, we again use the properties of UFDs. First, we can see that see that $b^2$ must divide $c_3^8$, and therefore $b$ divides $c_3$, such that $c_3 = p\, b$ for some polynomial $p$. At this point, there is the possibility that $p \equiv 0$ with generic $b$, greatly simplifying condition \eqref{eq:Z3-tuning_for_MP} to
\begin{align}\label{eq:Z3-tuningSimplified}
	b^4\,c_0^2 + 2\,b^2\,c_2 (c_1^2 - c_0\,c_2) + c_2^4 \equiv 0 \, .
\end{align}
Note that we have assumed that $b \neq 0 = c_3$ such that the generic elliptic curve is smooth, and dropped the overall powers of $b$ as the condition is trivially satisfied when $b=0$. In the same spirit as before, \eqref{eq:Z3-tuningSimplified} requires $b$ to divide $c_2$, such that there is a polynomial $t$ for which $c_2 = t\,b$. Assuming that neither $t$ nor $b$ is identically 0, we arrive at
\begin{align}
	2\,c_1^2\,t + b\,(c_0 - t^2)^2 = 0 \, .
\end{align}
This is formally of the form $A\,B = C\,D$, where the line bundle classes of individual terms on the left side does not match those on the right side. Hence, the generic solution must take the schematic form
\begin{align}\label{eq:Z3_tuning_intermediate_step}
\begin{split}
	& c_1^2 = q_1\,q_2 \, , \quad t = q_3\,q_4 \, , \\
	& b = -2\,q_1\,q_3 \, , \quad (c_0-t^2)^2 = q_2\,q_4 \, ,
\end{split}
\end{align}
with $(q_1, q_4)$ and $(q_2, q_3)$ being coprime.
But because $A = c_1^2$ is a complete square, $q_1$ and $q_2$ must combine into a square. As shown for instance in appendix A of \cite{Lawrie:2014uya}, this has for solution $q_i=r\eta^2_i,~i=1,2$, where $r$ is the factor common to $q_{1}$ and $q_2$ and $(\eta_1,\eta_2)$ are coprime.

Similarly, $(c_0-t^2)^2 = q_2\,q_4 = r\eta_2^2q_4$ implies that $rq_4$ must also be a square. However, as we have chosen a factorization such that $(q_1,q_4)$ are coprime, so must be $(r,q_4)$, as $r$ is a factor of $q_1$. 
Therefore, both $r$ and $q_4$ must be squares on their own, which in turn means that so must $q_1$ and $q_2$, and we deduce that the generic solution to \eqref{eq:Z3_tuning_intermediate_step} is
\begin{align}
	q_1 = a_1^2\, , \quad q_2 = a_2^2 \, , \quad q_4 = a_4^2 \,.
\end{align}
Relabeling $q_3$ as $a_3$, the section passes through the point of $\mathbb{Z}_3$ torsion -- under the simplification $p=0$ -- if the Weierstrass coefficients satisfy the following decomposition:
\begin{equation}
	 b = -2a_1^2a_3 \, , \quad c_0 = a_2a_4 + a_3^2a_4^4 \, , \quad c_1 = a_1a_2 \, , \quad c_2 = -2a_1^2a_3^2a_4^2 \, , \quad c_3 = 0 \, .
\end{equation}

A quick computation reveals that this solution indeed satisfies the tuning \eqref{eq:Z3-tuning_for_MP}. However, along the codimension 1 locus $\left\{ a_1=0 \right\}$, the Weierstrass functions vanish to order $\text{ord}(f,g,\Delta)=(4,6,12)$, indicating a non-minimal singularity type. In order to avoid the need to deal with such singularities, we impose that $a_1$ is a constant, which we rescale to 1 without loss of generality.
This identifies up to constants $b$ with $a_3$ and $c_1$ with $a_2$, and if we relabel $a_4 \rightarrow 2\,a$, the following decomposition is a solution to the $\bbZ_3$ torsional condition \eqref{eq:condition_for_Z3-torsion} of the Morrison--Park model without non-minimal singularities:
\begin{align}\label{eq:solution_Z3_MP}
	c_0 = 2\,c_1\,a + 4\,b^2\,a^4 \, , \quad c_2 = -2\,b^2\,a^2 \, , \quad c_3 = 0 \, .
\end{align}
The global validity of this solution is captured by the line bundle class of $a$. 
Comparing with the classes \eqref{tab:coefficients_classes_MP} of the other
coefficients, we find that this solution requires $a \in H^0(B , \CO(\beta + K_B))$.
Hence, the divisor $\beta + K_B$ must not be anti-effective. On $B = \bbP^2$ with hyperplane class $H$, this would imply $\beta = n\,H$ with $n \geq 3$.

\subsection[F-theory of the \texorpdfstring{$\mathbb{Z}_3$}{Z3} Torsional
Model and Higgsing to \texorpdfstring{$U(1)$}{U(1)}]{F-theory of the \boldmath{$\mathbb{Z}_3$}
Torsional Model and Higgsing to \boldmath{$U(1)$}}

Inserting the tuned solution \eqref{eq:solution_Z3_MP} into the Weierstrass
coefficients in \eqref{eqn:U1WS} yields
\begin{align}\label{eq:Weierstrass_functions_Z3-torsion}
\begin{split}
	& f_{\bbZ_3} = - \frac{2}{3} \, a\,b^2\,(8\,a^3\,b^2 + 3\,c_1) \, , \\
	& g_{\bbZ_3} = \frac{1}{108} \, b^2 \, (512 \, a^6 \, b^4 + 288 \, a^3 \, b^2 \, c_1 + 27\, c_1^2) \, , \\
	& \Delta_{\bbZ_3} = \frac{1}{16} \, b^4 \, c_1^3 \, (64 \, a^3 \, b^2 + 27 \, c_1) \, .
\end{split}
\end{align}
The vanishing orders indicate $I_3$ fibers along $\{c_1\}$ and type $IV$ fibers along $\{b\}$, both corresponding to an $\mathfrak{su}(3)$ gauge algebra.
Furthermore, there are two enhancement loci in codimension two, located at $\{c_1\} \cap \{b\}$ and $\{c_1\} \cap \{a\}$.
One easily verifies that at the second locus, the $I_3$ singularity over $\{c_1\}$ enhances to type $IV$, implying that there is no matter associated with that locus.
The other point sits at the intersection of the two $\mathfrak{su}(3)$ divisors and exhibits an $E_6$ singularity, and we therefore expect matter in the bifundamental representation.

To determine the multiplicity of the matter, we employ the cancellation
conditions of non-abelian anomalies. The multiplicities of adjoint matter is
again found using formula (\ref{intersecFormulaDiv}):
\begin{align}\label{eq:adjoint_multiplicity_su3s}
	\begin{split}
		& x_{\rm ad}^{\rm I_3} = 1+\frac{1}{2}  [c_1] \cdot ([c_1] + K_B) = 1 +
    \frac{1}{2}  (- K_B \cdot \beta + \beta^2) \, , \\
		& x_{\rm ad}^{\rm IV} = 1 + \frac{1}{2} \, [b] \cdot ([b] + K_B) = 1 +
    K_B^2 - \frac{1}{2} (-3 K_B \cdot \beta - \beta^2) \, .
	\end{split}
\end{align}
In a procedure similar to that of section \ref{sec:F-theory_on_Z2}, we find that all gauge anomalies are canceled if there is a complete hypermultiplet of bifundamental matter at each point in $\{c_1\} \cap \{b\}$, i.e.
\begin{align}\label{eq:bifund_multiplicity_su3s}
	x_{(\bf 3,3)} = [c_1] \cdot [b] = 2K_B^2 - K_B\cdot\beta - \beta^2 \, .
\end{align}
The absence of any fundamental matter as well as the existence of the by-construction $\bbZ_3$ torsional section indicate a non-trivial global gauge group structure of the F-theory compactification.
Indeed, one can verify that the section passes through the codimension one singularities of both the $I_3$ and type $IV$ fibers, proving that the global gauge group should be
\begin{align}
	\frac{SU(3)_{I_3} \times SU(3)_{IV}}{\bbZ_3} \, .
\end{align}

We can now discuss the breaking patterns of this model, in particular the breaking of the spectrum with gauge algebra
$\mathfrak{su}(3)_{I_3} \oplus \mathfrak{su}(3)_{IV}$ back to a
$U(1)$ model, similar to section \ref{sec:matching_spectrum} in case of the
$\bbZ_2$ torsional model.
One possible way of doing so is to first break both $\mathfrak{su}(3)$ factors
to their Cartan subalgebra by giving a vev to an adjoint hypermultiplet of
each factor, resulting in a $U(1)_{\rm I_3}^2 \times U(1)_{\rm IV}^2$ gauge
group. Each Cartan subalgebra can be further broken with a charged state
arising from the remaining adjoint hypermultiplets. This yields a $U(1)_{\rm
I_3} \times U(1)_{\rm IV}$ gauge group, where each $U(1)$ is in the Cartan of
one of the $\mathfrak{su}(3)$ factors. At this step the bifundamental
hypermultiplets always decomposes as 
\begin{align}\label{eq:su3_bifund_remnants}
	({\bf 3, 3}) \longrightarrow 4 \times (1,1) \oplus 2 \times (2,-1) \oplus 2\times (-1,2) \oplus (2,2) \, .
\end{align}

The only two possibilities to further Higgs to a $U(1)$ model with just charge
$1$ and $2$ singlets is to give a vev either to the states with charges $(2,-1)$ or $(-1,2)$.
If we higgs with the states $(2,-1)$ under $U(1)_{I_3} \times U(1)_{IV}$, then
the adjoints of $\mathfrak{su}(3)_{I_3}$ yield charge $2$ singlets, and those of
$\mathfrak{su}(3)_{IV}$ yield charge $1$, and vice versa for Higgsing with
$(-1,2)$ states.  Note that this discussion is purely group theoretical and
the subscripts are merely labels.  The geometry, however, differentiates the
two factors as the multiplicities are different, and we therefore do not
expect the two Higgs chains to be equivalent.

Based on the multiplicities \eqref{eq:adjoint_multiplicity_su3s} and \eqref{eq:bifund_multiplicity_su3s}, we find that Higgsing with charge $(-1,2)$ singlets produces a Morrison--Park model characterized by the class $\beta$, i.e., it is the $U(1)$ model with which we started the tuning process \eqref{eq:solution_Z3_MP}.
The other possible Higgsing, with $(2,-1)$ states, yields the following multiplicities of charged singlets:
\begin{align}
	\begin{split}
		& x_1 = 16\, K_B^2 - 4 \beta^2 \, , \\
		& x_2 = 2\,K_B^2 - 3K_B\cdot \beta + \beta^2 \, .
	\end{split}
\end{align}
Formally, this looks like the spectrum of a Morrison--Park $U(1)$ model with twisting line bundle class $\tilde{\beta} = -K_B - \beta$.
However, as we discussed just below \eqref{eq:solution_Z3_MP}, the class $\beta + K_B$ must not be anti-effective.
Therefore, the putative class $\tilde{\beta}$ does not give rise to a well-defined Morrison--Park model.
Hence, the second Higgsing chain is geometrically obstructed.
We can back-track this obstruction explicitly to the fact that the tuning \eqref{eq:solution_Z3_MP} does not allow for any other identification of the Morrison--Park coefficients in terms of the polynomials defining the $\bbZ_3$ tuned model \eqref{eq:Weierstrass_functions_Z3-torsion}.

\bibliography{FTheory}{}
\bibliographystyle{JHEP} 

\end{document}